\documentclass[]{aa}
\usepackage{natbib}
\bibpunct{(}{)}{;}{a}{}{,} 
\usepackage[varg]{txfonts}
\usepackage[squaren, thinspace,thickqspace, textstyle]{SIunits}
\usepackage{dblfloatfix}
\usepackage{subfig}
\usepackage{textcomp}
\usepackage{color}
\usepackage{bm}
\usepackage{multirow}
\usepackage{booktabs}

\newcommand{\cmmnt}[1]{}
\newcommand{\mockalph}[1]{}

\begin{document}

\title{The CARMENES search for exoplanets around M dwarfs}

\subtitle{A Neptune-mass planet traversing the habitable zone around HD\,180617}

 \author{A.~Kaminski\inst{\ref{inst:zah_lsw}}
     \and T.~Trifonov\inst{\ref{inst:mpia}}
     \and J.~A.~Caballero\inst{\ref{inst:CSIC-INTA}}
     \and A.~Quirrenbach\inst{\ref{inst:zah_lsw}}
     \and I.~Ribas\inst{\ref{inst:CSIC-ICE},\ref{inst:IEEC}}  
     \and A.~Reiners\inst{\ref{inst:iag_goett}}
     \and P.~J.~Amado\inst{\ref{inst:IAA-CSIC}} 
     \and M.~Zechmeister\inst{\ref{inst:iag_goett}}
     \and S.~Dreizler\inst{\ref{inst:iag_goett}}
     \and M.~Perger\inst{\ref{inst:CSIC-ICE},\ref{inst:IEEC}}
     \and L.~Tal-Or\inst{\ref{inst:TelAviv},\ref{inst:iag_goett}}
     \and X.~Bonfils\inst{\ref{inst:grenoble}}
     \and M.~Mayor\inst{\ref{inst:obs_geneve}}
     \and N.~Astudillo-Defru\inst{\ref{inst:uni_conc}}
     \and F.~F.~Bauer\inst{\ref{inst:IAA-CSIC},\ref{inst:iag_goett}}
     \and V.~J.~S.~B\'ejar\inst{\ref{inst:iac_ten},\ref{inst:da_ten}}
     \and C.~Cifuentes\inst{\ref{inst:dep_aca_madrid}, \ref{inst:CSIC-INTA}}
     \and J.~Colom\'e\inst{\ref{inst:CSIC-ICE},\ref{inst:IEEC}}
     \and M.~Cort\'es-Contreras\inst{\ref{inst:CSIC-INTA}}
     \and X.~Delfosse\inst{\ref{inst:grenoble}}
     \and E.~D\'iez-Alonso\inst{\ref{inst:dep_aca_madrid}}    
     \and T.~Forveille\inst{\ref{inst:grenoble}}
     \and E.~W.~Guenther\inst{\ref{inst:th_lsw}}
     \and A.~P.~Hatzes\inst{\ref{inst:th_lsw}}
     \and Th.~Henning\inst{\ref{inst:mpia}}
     \and S.~V.~Jeffers\inst{\ref{inst:iag_goett}}
     \and M.~K\"urster\inst{\ref{inst:mpia}}
     \and M.~Lafarga\inst{\ref{inst:CSIC-ICE},\ref{inst:IEEC}}
     \and R.~Luque\inst{\ref{inst:iac_ten},\ref{inst:da_ten},\ref{inst:zah_lsw}}
     \and H.~Mandel\inst{\ref{inst:zah_lsw}}
     \and D.~Montes\inst{\ref{inst:dep_aca_madrid}} 
     \and J.~C.~Morales\inst{\ref{inst:CSIC-ICE},\ref{inst:IEEC}}
     \and V.~M.~Passegger\inst{\ref{inst:ham_sw}}
     \and S.~Pedraz\inst{\ref{inst:CSIC-MPG}} 
     \and S.~Reffert\inst{\ref{inst:zah_lsw}}
     \and S.~Sadegi\inst{\ref{inst:zah_lsw}}  
     \and A.~Schweitzer\inst{\ref{inst:ham_sw}}
     \and W.~Seifert\inst{\ref{inst:zah_lsw}}
     \and O.~Stahl\inst{\ref{inst:zah_lsw}}  
     \and S.~Udry\inst{\ref{inst:obs_geneve}}
}
  \institute{ Landessternwarte, Zentrum f\"ur Astronomie der Universt\"at Heidelberg,
              K\"onigstuhl 12, D-69117 Heidelberg, Germany\label{inst:zah_lsw}\\
              \email{A.Kaminski@lsw.uni-heidelberg.de}
         \and Max-Planck-Institut f\"ur Astronomie,
              K\"onigstuhl 17, D-69117 Heidelberg, Germany\label{inst:mpia}
         \and Centro de Astrobiolog\'ia (CSIC-INTA), ESAC campus, Camino Bajo del Castillo s/n, 
              E-28692 Villanueva de la Ca\~nada, Madrid, Spain\label{inst:CSIC-INTA}
         \and Institut de Ci\`encies de l’Espai (ICE, CSIC), Campus UAB, c/ de Can Magrans s/n, 
              E-08193 Bellaterra, Barcelona, Spain\label{inst:CSIC-ICE}
         \and Institut d’Estudis Espacials de Catalunya (IEEC), E-08034 Barcelona, Spain \label{inst:IEEC}
         \and Institut f\"ur Astrophysik, Georg-August-Universit\"at,
              Friedrich-Hund-Platz 1, D-37077 G\"ottingen, Germany\label{inst:iag_goett}
         \and Instituto de Astrof\'isica de Andaluc\'ia (IAA-CSIC), Glorieta de la Astronom\'ia s/n,
              E-18008 Granada, Spain\label{inst:IAA-CSIC}
         \and School of Geosciences, Raymond and Beverly Sackler Faculty of Exact Sciences, 
              Tel Aviv University, Tel Aviv 6997801, Israel\label{inst:TelAviv}
         \and Universit\'e Grenoble Alpes, CNRS, IPAG, 38000 Grenoble, France\label{inst:grenoble}
         \and Observatoire de Gen\`eve, Universit\'e de Gen\`eve, 51 ch. des Maillettes, 1290 Sauverny, Switzerland\label{inst:obs_geneve}
         \and Departamento de Astronom\'ia, Universidad de Concepci\'on, Casilla 160-C, Concepci\'on, Chile\label{inst:uni_conc}
         \and Instituto de Astrof\'isica de Canarias, V\'ia L\'actea s/n, 38205 La Laguna,
              Tenerife, Spain\label{inst:iac_ten}
         \and Departamento de Astrof\'isica, Universidad de La Laguna,
              E-38206 La Laguna, Tenerife, Spain\label{inst:da_ten}
         \and Departamento de Astrof\'isica y Ciencias de la Atm\'osfera,
              Facultad de Ciencias Físicas, Universidad Complutense de Madrid,
              E-28040 Madrid, Spain\label{inst:dep_aca_madrid}
         \and Th\"uringer Landessternwarte Tautenburg, Sternwarte 5, D-07778 Tautenburg, Germany\label{inst:th_lsw}
         \and Hamburger Sternwarte, Gojenbergsweg 112, D-21029 Hamburg, Germany\label{inst:ham_sw}
         \and Centro Astron\'omico Hispano-Alem\'an (CSIC-MPG),
              Observatorio Astron\'omico de Calar Alto,
              Sierra de los Filabres, E-04550 G\'ergal, Almer\'ia, Spain\label{inst:CSIC-MPG}
              }

\date{Received 3 May 2018 / Accepted DD MM 2018}

\abstract{Despite their activity, low-mass stars are of particular importance for the search of exoplanets by the means of Doppler spectroscopy, as planets with lower masses become detectable.
We report on the discovery of a planetary companion around HD\,180617, a bright ($J = 5.58$\,mag), low-mass ($M = 0.45$\,M$_\odot$) star of spectral type M2.5\,V. The star, located at a distance of 5.9\,pc, is the primary of the high proper motion binary system
containing vB\,10, a star with one of the lowest masses known in most of the twentieth century. Our analysis is based on new radial velocity (RV) measurements 
made at red-optical wavelengths provided 
by the high-precision spectrograph CARMENES, which was designed to carry out a survey for Earth-like planets around M dwarfs. The available CARMENES data are augmented by archival Doppler measurements from 
HIRES and HARPS. Altogether, the RVs span more than 16 years.
The modeling of the RV variations, with a semi-amplitude of $K = 2.85_{-0.25}^{+0.16}\,\metrepersecondnp$, yields a Neptune-like planet with a minimum mass of $12.2_{-1.4}^{+1.0}$~M$_\oplus$ on a $105.90_{-0.10}^{+0.09}$~d circumprimary orbit, which is partly located
in the host star's habitable zone. The analysis of time series of common activity indicators does not show any dependence on the detected RV signal. The discovery of HD\,180617\,b not only adds information to a currently hardly filled region of the mass-period
diagram of exoplanets around M dwarfs, but the investigated system becomes the third known binary consisting of M dwarfs and hosting an exoplanet in an S-type configuration. Its proximity makes it an attractive candidate for future studies.
}

 
\authorrunning{A.~Kaminski et al.}
\titlerunning{Neptune-like planet around HD\,180617}

\maketitle 

\section{Introduction}\label{sec:intro}
The search for exoplanets by means of Doppler spectroscopy has significantly advanced since the first confirmed discoveries at the end of the last century \citep[e.g.,][]{Queloz_1995}.
The development in instrumentation within this field led to detection limits on the order of meters per seconds, as has been demonstrated, for example, by HIRES at the Keck Observatory  \citep{Vogt_1994},
HARPS at the La Silla Observatory \citep{Mayor_2003}, HARPS-N at El Roque de Los Muchachos Observatory \citep{Consentino_2014}, or the Automated Planet Finder at the Lick Observatory  \citep{Rado_2014}.\\ \indent
Still, the choice of the parent stars is crucial. It allows pushing the boundaries 
with respect to the exoplanet parameter space, and exoplanet programs targeting M dwarfs have become of major interest (\citealp[e.g.,][]{Char_2008}; \citealp{Zechmeister_2009a}; \citealp{Bon_2013}).
Because of their low masses, M dwarfs are particularly suitable for Doppler 
surveys, since the semi-amplitude of a radial velocity (RV) signal induced by a companion increases with decreasing stellar mass. Thus, lower mass planets are more easily detectable around low-mass stars. 
M dwarfs have luminosities between $10^{-4}$ and $10^{-1} \mathrm{L}_\odot$ and produce significantly less flux than their more massive counterparts. Consequently, their habitable zones (HZs) are located
closer to their host stars at distances of about 0.05 to 0.4\,au (\citealp[e.g.,][]{Joshi_1997}; \citealp{Tarter_2007}; \citealp{Kopp_2014}; \citealp{Dress_2015}).\\ \indent
However, the intrinsic stellar activity of M dwarfs can hinder the search for exoplanets around these stars. Signatures of stellar activity such as active regions on the stellar surface and their cyclic
variability can mimic radial velocity signals that have previously been misinterpreted as arising from a planetary companion. 
Since the strength of such activity-related signals has been shown to be wavelength dependent \citep[e.g.,][]{Desort_2007, Reiners_2010}, a wide 
wavelength coverage in RV measurements can help distinguish planetary signals from those due to activity \citep[e.g.,][]{Sarkis_2018}.\\ \indent
The high-resolution fiber-fed spectrograph CARMENES\footnote{\tt http://carmenes.caha.es}, 
which is installed at the 3.5~m  telescope at the Calar Alto Observatory in Spain, was specifically designed with a broad wavelength coverage. 
It consists of two spectrograph channels, which together cover wavelengths from 520 to
1710~nm, with a resolution of 94\,600 in the visual and 80\,500 in the near-infrared channel \citep{Quirrenbach_2014, Quirrenbach_2016}.
The instrument has been operating since January 2016 and is performing a search for exoplanets around M dwarfs. More than 300 targets that have been selected from the CARMENES input catalog 
Carmencita \citep{Caballero_2016b, Reiners_2018b} are regularly monitored for this purpose.
The capability of the visual channel of achieving an RV precision of 1--2~$\metrepersecondnp$ has been demonstrated by \citet{Seifert_2016}, \citet{Trifonov_2018}, and the first CARMENES exoplanet discovery by \citet{Reiners_2018a}.  \\ \indent
In this paper we analyze the RV data of HD\,180617, one of the M dwarfs monitored by CARMENES. The measurements
indicate the presence of a planet with a minimum mass comparable to Neptune on an orbit partly located within the habitable 
zone of the host.
In Section~\ref{sec:stars} we characterize the observed star, followed by a short description of the RV data compilation in Section~\ref{sec:data}. 
The results from the RV analysis are presented in 
Section~\ref{sec:RV_ana}, and we conclude in Section~\ref{sec:conclusions}.

\section{Host star HD\,180617 (GJ 752\,A)}\label{sec:stars}
The star HD\,180617 is a well-investigated early-M dwarf.
Discovered from K\"onigstuhl almost one century ago by \cite{Wolf_1919}, HD\,180617 is known particularly for being the primary of the common (high) proper motion pair that also contains the M8.0\,V dwarf \object{vB\,10} (V1298\,Aql).
The pair, first reported as a binary by \citet{van_Bies_1944}, received the code LDS~6334 in the Washington Double Star catalog \citep{Mason_2001}, and is separated by about 76 arcseconds.
With its mass near the hydrogen-burning limit at 0.07\,M$_\odot$, vB\,10 was the least massive star known for almost four decades \citep{van_Bies_1944, Herb_1956, Kum_1964, Kirk_1991}.
\citet{Prav_2009} proposed that vB\,10 might host a giant planet, which made ``vB\,10\,b'' the first putative exoplanet around an ultracool dwarf and the first such object discovered astrometrically.
However, this hypothesis was soon after refuted by means of precise radial velocity and astrometric measurements \citep{Bean_2010, Anglada_2010, Laz_2011}.  \\ \indent
   \begin{table}
      \caption{Basic information on the host star.} %
         \label{table.stars}
\centering
         \begin{tabular}{l l r}
            \hline
            \hline
            \noalign{\smallskip}
                & \object{HD\,180617}         & Ref.$^{a}$ \\
            \hline
            \noalign{\smallskip}
Karmn$^{b}$                 & J19169+051N                 &     \\
Wolf                 & 1055                 &     \\
GJ                 & 752\,A                 &        \\
BD                & +04 4048                &        \\
Var. name                       & V1428\,Aql                            &               \\
\noalign{\smallskip}
Sp. type            & M2.5\,V                & AF15    \\
$G$ [mag]            & 8.0976$\pm$0.011            & {\em Gaia}     \\
$J$ [mag]            & 5.583$\pm$0.030            & 2MASS     \\
$d$ [pc]            & 5.9116$\pm$0.018            & {\em Gaia} \\
$\mu_\alpha \cos{\delta}$ [mas\,a$^{-1}$]       & --579.043$\pm$0.088$^{c}$     & {\em Gaia} \\
$\mu_\delta$ [mas\,a$^{-1}$]            & --1332.743$\pm$0.081$^{c}$    & {\em Gaia} \\
$V_r$ [km\,s$^{-1}$]        & +35.678                & Rei18        \\
$U$ [km\,s$^{-1}$]      & +53.2         & This work     \\
$V$ [km\,s$^{-1}$]      & --7.6         & This work     \\
$W$ [km\,s$^{-1}$]      & --5.0         & This work     \\
$v \sin{i}$ [km\,s$^{-1}$]    & $<$2 & Rei18        \\
$T_{\rm eff}$ [K]        & 3557$\pm$51                & Pas18        \\
$\log{g}$             & 4.86$\pm$0.07                & Pas18        \\
{[Fe/H]}                 & 0.00$\pm$0.16            & Pas18        \\
$L$ [L$_\odot$]    & 0.0326$\pm$0.0004                & This work     \\
$R$ [R$_\odot$]            & 0.453$\pm$0.019            & This work    \\
$M_\star$ [M$_\odot$]        & 0.45$\pm$0.04 & This work \\
pEW(H$\alpha$) [{\AA}]        & +0.3$\pm$0.1 & Jef18 \\
$\log{R'_\mathrm{HK}}$ & -5.071$\pm$0.071 & Ast17\\
$P_{\rm rot}$ [d]        & 46.04$\pm$0.20            & DA17\\
Kinematic pop.            & Thin disc                & CC16        \\
        \noalign{\smallskip}
            \hline
         \end{tabular}
\begin{list}{}{}
\item[$^{a}$] References --
2MASS: \citet{Skru_2006};
AF15: \citet{Alo_2015};
Ast17: \citet{Astudillo_2017};
CC16: \citet{Cor_2016_phd};
DA17: \citet{Diez_2017}; 
{\em Gaia}: \citet{Gaia_2018};
Jef18: \citet{Jef_2018}; 
Pas18: \citet{Pas_2018};  
Rei18: \citet{Reiners_2018b}. 
\item[$^{b}$] Carmencita identifier \citep{Caballero_2016b}.
\item[$^{c}$] Proper motions from the Tycho-{\em Gaia} Astrometric Solution \citep{Gaia_2016} have slightly smaller uncertainties, but for homogeneity, we use those from {\em Gaia} DR2. 
\end{list}
   \end{table}
The actual exoplanet host is the primary star HD~180617. 
Although it displays Ca~{\sc ii} H\&K in emission \citep{Lipp_1952} and is listed as a flaring star by Simbad, it is rather inactive.
In spite of its M2.5\,V spectral type, it displays H$\alpha$ in absorption \citep{Jef_2018} and very faint X-ray emission \citep{Gonz_2014_MSc}.
With parallactic distance and proper motions from the {\em Gaia} Data Release 2 \citep[DR2;][]{Gaia_2018} and radial velocity from \cite{Reiners_2018b}, we computed Galactocentric space velocities $UVW$ as in \cite{Mon_2001}, which within errors
agree with \cite{Delfosse_1998},
and we conclude as \cite{Cor_2016_phd} did that it is kinematically separated from the young stellar population in the thin disk.
The star exhibits a solar-like metallicity \citep{Pas_2018}, and its most probable age therefore lies in the wide range between 1 and 10\,Gyr. The extremely young age derived by \citet{Tetz_2011} for HD~180617 is apparently incorrect.  \\ \indent
The basic information on the star is given in Table~\ref{table.stars}.
The tabulated values are in general consistent with previous determinations, when available
(e.g., radial velocities by \citealt{Nid_2002}, \citealt{Nord_2004}, and \citealt{Soub_2013};
spectral type by \citealt{Joy_1974} and \citealt{Hen_2002};
photospheric parameters by \citealt{Soub_2008}, \citealt{Roj_2012}, and \citealt{Gaid_2014}).  \\ \indent
To determine new stellar parameters, we first collected broadband photometry from several surveys covering the whole spectral energy distribution (SED) of all CARMENES GTO targets \citep{Caballero_2016b}.
In order to determine the stellar luminosity $L$, we used the Virtual Observatory SED Analyzer \citep{Bayo_2008}, the {\em Gaia} parallactic distance, and photometry from the following catalogs:
SDSS \citep{Ahn_2012},
UCAC4 \citep{Zach_2013},
Tycho-2 \citep{Hog_2000},
{\it Gaia} DR2 \citep{Gaia_2018},
CMC15 \citep{Mui_2014},
2MASS \citep{Skru_2006},
and {\it AllWISE} \citep{Cut_2013}. 
In the {\it GALEX} passbands \citep{Bian_2011} and SDSS $u$, the M-dwarf photometry is dominated by chromospheric emission.
Using this luminosity together with the spectroscopic $T_{\rm eff}$ from \citet{Pas_2018} and the Stefan-Boltzmann law, we calculated the radius $R$.
To derive the stellar mass $M,$ we did not use the surface gravity, as we find the large uncertainty that is introduced by the error propagation unacceptable.
Instead, we applied the empirical linear $M_\star$-$R_\star$ relation determined by \citet{Schweitzer_prep},
which is similar to other determinations in the literature (e.g., \citealt{Torr_2013} and references therein).
Based on data from \citeauthor{Poj_2002} (\citeyear{Poj_2002}; ASAS), we list a rotational period of $P_{\rm rot}$ = 46.04$\pm$0.20\,d measured by \citeauthor{Diez_2017} (\citeyear{Diez_2017}; DA17).
The reported rotational period was determined by standard means of signal search with the Lomb-Scargle periodogram of
the time series of 389 observations, spread over eight years. It is almost identical to the period derived by \citeauthor{Sua_2016} (\citeyear{Sua_2016}; SM16), as
both determinations are based on the same data set. There are differences between amplitudes (8\,mmag in DA17, 4.5\,mmag in SM16) and false-alarm probabilities (FAP; 2.0\,\% in DA17; $<$ 0.1\,\% in SM16), however.
The period also agrees with the one estimated from the stellar $\log{R'_{\rm HK}}$ of --5.07, $P_{\rm HK}$ = 47$\pm$4\,d \citep{Astudillo_2017}.\\ \indent
More than 700 citations are collected for HD\,180617 and vB\,10 together, but numerous misidentifications of the secondary still abound in public catalogs because of its low Galactic latitude ($b \approx$ --3\,deg) and the resulting high stellar density in the area.
Furthermore, \citet{Cor_2014} noted that WDS tabulated only three epochs from 1942 to 1999, and the last epoch (2MASS) was incorrect.
In Table~\ref{table.astrometry_V1428} we provide angular separations, position angles, epochs, and the
sources of the relative astrometry determined by us.
With the 33-year time baseline, and in spite of the saturation of the primary in
the photographic plates (particularly from the Quick-$V$ Northern survey), we
double the number of published astrometric visits for this system and confirm
that there is no measurable orbital variation. The pair is separated by $\rho = 75.20\pm0.22$\,arcsec, which translates into a projected physical separation of $444.6\pm1.3$\,au.
The expected orbital period is about 10$^4$ years \citep{Cor_2014}.  \\ \indent
Using CARMENES data, \citet{Reiners_2018b} tabulated the first precise determination of the absolute radial velocity of vB\,10 to $V_r$ = +35.699\,km\,s$^{-1}$.
This is almost identical to that of HD\,180617 (see Table~\ref{table.stars}), as expected from their common proper motion and distance ({\em Gaia} DR2 tabulates $d$ = 5.918$\pm$0.005\,pc for vB~10).
In spite of numerous deep searches \citep{Opp_2001, Car_2005, Jod_2013, Ward_2015}, no additional stellar or substellar member in the system could be confirmed to date.
\begin{table}
      \caption{Astrometric measurements of the wide pair HD\,180617~+~vB\,10 (WDS~19169+0510, LDS~6334)$^{a}$.}
         \label{table.astrometry_V1428}
\centering
         \begin{tabular}{ccc ll}
            \hline
            \hline
            \noalign{\smallskip}
$\rho$        & $\theta$    & Epoch        & Band & Origin    \\
$[\mathrm{arcsec}]$    & $[\mathrm{deg}]$        &     \\
            \noalign{\smallskip}
            \hline
            \noalign{\smallskip}
74.8$\pm$0.4    & 152.0$\pm$0.2    & 1982.562     & Q$V$     & Quick-$V$ \\&&&&Northern \\
75.6$\pm$0.4    & 151.9$\pm$0.2    & 1983.673     & Q$V$    & Quick-$V$ \\&&&&Northern \\
75.2$\pm$0.2    & 152.5$\pm$0.2    & 1992.585     & $R_F$    & POSS-II \\&&&&Red    \\
75.1$\pm$0.2    & 152.0$\pm$0.2    & 1994.368     & $I_N$    & POSS-II \\&&&&Infrared \\
75.1$\pm$0.4    & 152.0$\pm$0.2    & 1995.621     & $B_J$ & POSS-II \\&&&&Blue    \\
75.22$\pm$0.06  & 152.0$\pm$0.1    & 1999.578      & $JHK_s$ & 2MASS    \\
75.13$\pm$0.08  & 152.0$\pm$0.1    & 2004.519     & $r'$    & CMC15    \\
75.19$\pm$0.07  & 152.3$\pm$0.1    & 2010.521    & $W1W2$ & {\em AllWISE} \\
75.8$\pm$0.4    & 152.1$\pm$0.4    & 2012.755    & $R$    & CC14    \\
75.4993$\pm$0.0016   & 152.491$\pm$0.002 & 2015.500    & $G$    & {\em Gaia}    \\
        \noalign{\smallskip}
            \hline
         \end{tabular}
\begin{list}{}{}
\item[$^{a}$] Compiled from all-sky surveys
(2MASS: \citealt{Skru_2006};
CMC15: \citealt{Mui_2014};
{\em AllWISE}: \citealt{Cut_2013}; 
{\em Gaia}: \citealt{Gaia_2018}),
CC14: \citealt{Cor_2014},
or measured by us on digitizations of photographic plates
(Digital Sky Survey digitizations of the Quick-$V$ Northern survey used for the
{\em Hubble Space Telescope} Guide Star Catalogue 1,
and SuperCOSMOS digitizations of blue, red, and infrared Palomar Observatory Sky
Survey).
We also investigated the latest data releases of the Astrographic Catalogue,
{\em AKARI}, {\em Gaia} DR1, and PanSTARRS, among others.
\end{list}
   \end{table}

\section{Radial velocity data}\label{sec:data}
We initially considered the RVs from both CARMENES channels for HD\,180617. However, \citet{Reiners_2018b} showed that the 
intrinsic precision (i.e., RV content) of the near-infrared channel velocities is about four times lower than that of the visual channel for 
targets of M2--3 spectral type. Given the low amplitude of the RV signal and the small contribution of the near-infrared RVs to improving 
the model parameters, we decided to use only the CARMENES visual channel data for our analysis. A complete characterization of the 
near-infrared channel and its performance will be carried out in a broader context of the survey. It is work in progress and beyond the scope of this particular publication.\\ \indent
All exposures used for this work comprise the spectral information of both the stellar target and the calibration source, which is a temperature-stabilized Fabry-P\'erot
etalon \citep{Schaefer_2012} that monitors any instrument drifts during the night. These drifts are typically below 10\,$\metrepersecondnp$ and are determined with a precision below 1\,$\metrepersecondnp$.
Together with the standard dusk and dawn calibration sequences, which contain exposures of U-Ar, U-Ne, and Th-Ne hollow cathode lamps, 
an accurate wavelength reference is provided for each observation. The wavelength solution, described in \citet{Bauer_2015}, is encapsulated into the standard
data processing, carried out by the CARACAL pipeline \citep{Caballero_2016a}. The software also performs standard corrections such as bias subtraction and flat-fielding. \\ \indent
From the order-by-order extracted spectra, the radial velocities are then derived by the RV pipeline SERVAL \citep[SpEctrum Radial Velocity AnaLyser;][]{Zechmeister_2017}. 
The RV computation is based on least-squares fitting, where the RVs are determined against a template with a high signal-to-noise ratio that is constructed by coadding all available spectra of the target.
They are corrected for barycentric motion \citep{Wright_2014}, as well as for secular 
acceleration, which is important for stars with high proper motions \citep{Zechmeister_2009a}.\\ \indent
In addition, a nightly zero-point (NZP) correction was applied in order to achieve the highest RV precision at the $1\metrepersecondnp$ level. For each night, the NZPs were determined from 
RV measurements of the survey's ``RV-quiet'' stars, which in general show only little RV variability. This correction, whose values for the observation nights of HD\,180617 were on average about 3\,$\metrepersecondnp$, 
was verified and described in more detail by \citet{Trifonov_2018}.\\ \indent
We supplemented the CARMENES RV measurements by data available from the HARPS spectrograph in Chile and the HIRES spectrograph in Hawai'i. Consequently, our final Keplerian models were applied to the 
combined data sets.
The HIRES data were taken from \citet{Butler_2017}, while the HARPS RVs were calculated by us using the ESO archival spectra that were reprocessed by SERVAL.
All the Doppler measurements used within this work are listed in Table~\ref{table:V1428_RVs_a}.
We combined 124 CARMENES RV measurements taken over 622 days, 138 HARPS measurements over 4470 days, and 158 HIRES RVs spanning 4849 d. 
One data point (at JD = 2457258.622) was removed from the HARPS data set, as it is an obvious outlier with an offset of more than $10\,\metrepersecondnp$. 
The full data set of 421 measurements spans 6037 days. There is some overlap in time between the HIRES and HARPS measurements, and also between the HARPS and CARMENES RVs.
The  typical RV precisions of the instruments are estimated from the medians of the RV uncertainties at 2.10\,ms$^{-1}$ for HIRES, 0.76\,ms$^{-1}$ for HARPS, and 1.42\,ms$^{-1}$ for CARMENES. Compared to the HARPS RVs determined by SERVAL, 
 those determined by the standard HARPS-DRS pipeline in the case of HD\,180617 show lower errors with a median of 0.43\,ms$^{-1}$. This is most likely explained by the different error treatment of the two distinct software packages, and does not in general
imply a difference in accuracy. We found that our results (see Sect.~\ref{ssec:RV_model}) agree very well with those obtained together with the HARPS DRS RVs.

\section{Analysis and results}\label{sec:RV_ana}
\subsection{Periodogram analysis}\label{ssec:RV_GLS}
\begin{figure}[]
\resizebox{\hsize}{!}{\includegraphics{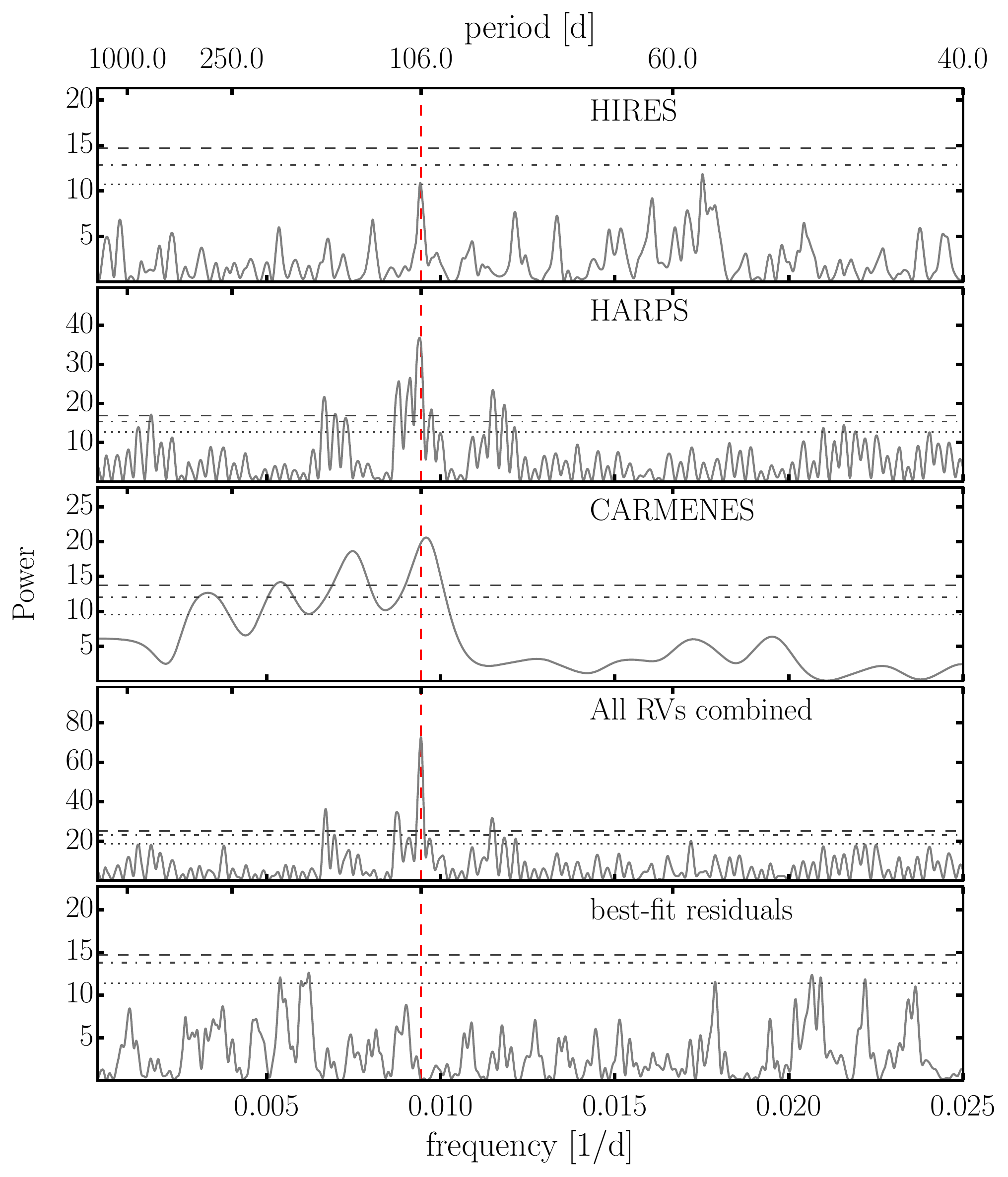}}

\caption{GLS periodograms of the RV data for HD\,180617. The upper three panels represent the individual instruments, followed by the combined data set. The lowermost panel shows the periodogram of the residuals
of the best-fit solution.
The horizontal lines (dotted, dash-dotted, and dashed) illustrate FAP levels of 10\%, 1\%, and 0.1\%. The best-fit period at $P_b = $ 105.9\,d is highlighted by the vertical red line. 
}
\label{fig:gls_RV_V1428}
\end{figure}

In Fig.~\ref{fig:gls_RV_V1428} we provide generalized Lomb-Scargle (GLS) periodograms \citep{Zechmeister_2009b} for the individual data sets (CARMENES, HARPS, and HIRES) as well as 
the combined data. Three levels of FAPs at 0.1\%, 1\% and 10\% are also depicted. They were assessed over the whole frequency range by means of
bootstrapping, for which 1000 randomly reordered time series were created from the original data.
The most significant peak in the CARMENES periodogram appears at a period of $P \approx $ 104\,d ($f \approx$ 0.00961\,d$^{-1}$) with an FAP below 0.1\%. 
Three other signals are visible at periods of 
around 134, 184 (FAPs below 0.1\%), and 300 days (FAP below 1\%), which translates into frequencies of 0.00744\,d$^{-1}$, 0.00543\,d$^{-1}$, and 0.00334\,d$^{-1}$. Since the four signals seem
equidistant with $\Delta f \approx 0.00209$\,d$^{-1}$ in frequency space, and the sampling function of the CARMENES RVs shows a prominent peak close to $\Delta f$ at around
$f \approx 0.00233$\,d$^{-1}$ (428\,d), one might assume that the three peaks with longer periods are aliases of the most significant signal. However, when the dominant peak at 104\,d is subtracted from
the data, only the signal at 134\,d vanishes. The other two remain, and the peak at around 300\,d becomes the most significant. Subtracting a simultaneous fit to the two signals at 104\,d and 300\,d from the data 
eliminates all of the four peaks. The same applies to a simultaneous fit to the signals at 104\,d and 184\,d. We therefore  conclude that the signal at 134\,d is an alias of the 104\,d peak and that the signals at 184\,d and at about 300\,d
are also correlated. It is unclear which of the latter two is the true signal, which an alias, and where they come from. The 300\,d signal is only visible in the periodogram of the CARMENES data, 
although HARPS shows a signal with an FAP below 1\% at around 600\,d. While the time span of the CARMENES measurements
is currently too short for properly sampling such a long period, the 300\,d peak could be a harmonic of it.
On the other hand, the peak at 184\,d, which is apparent also in the model residuals, is close to four times the rotation period of the star and might be correlated to some activity signals we find (see below).
In addition to the most significant peak at $P \approx $ 106\,d with a power of around 38 and the peak at around $P \approx $ 600\,d, the HARPS periodogram shows further signals that can be attributed to modulations that are due to the sampling function. The latter
shows dominant peaks at $f_1 \approx 0.00271$\,d$^{-1}$ (369\,d) and at $f_2 \approx 0.00033$\,d$^{-1}$ (3005\,d). Consequently, aliases appear at about $f_a=f \pm f_s$ (0.00668\,d$^{-1}$ and 0.01210\,d$^{-1}$). These aliases, as well as the main signal, come in packages,
where the peaks are separated by $\Delta f \approx f_2$.
The HIRES periodogram does not exhibit any significant peak. There is subtle power (FAP around 10\%) at 106\,d, as well as at around 0.01754\,d$^{-1}$ (57\,d). The latter is not easily explained by the sampling function, which shows dominant
peaks at periods of one year and one month only. Therefore its nature remains unclear.\\ \indent
In the GLS periodogram of the combined data set, the signal in question is highly significant (FAP $<$ 0.002\%), with a refined period value of around 106\,d. Since all additional peaks of significance are understood, mainly due to the sampling of the HARPS time series, 
 we focused on the dominant peak alone for the further modeling.

\subsection{Keplerian modeling}\label{ssec:RV_model}
\begin{figure*}[]
\centering
\includegraphics[width=18cm]{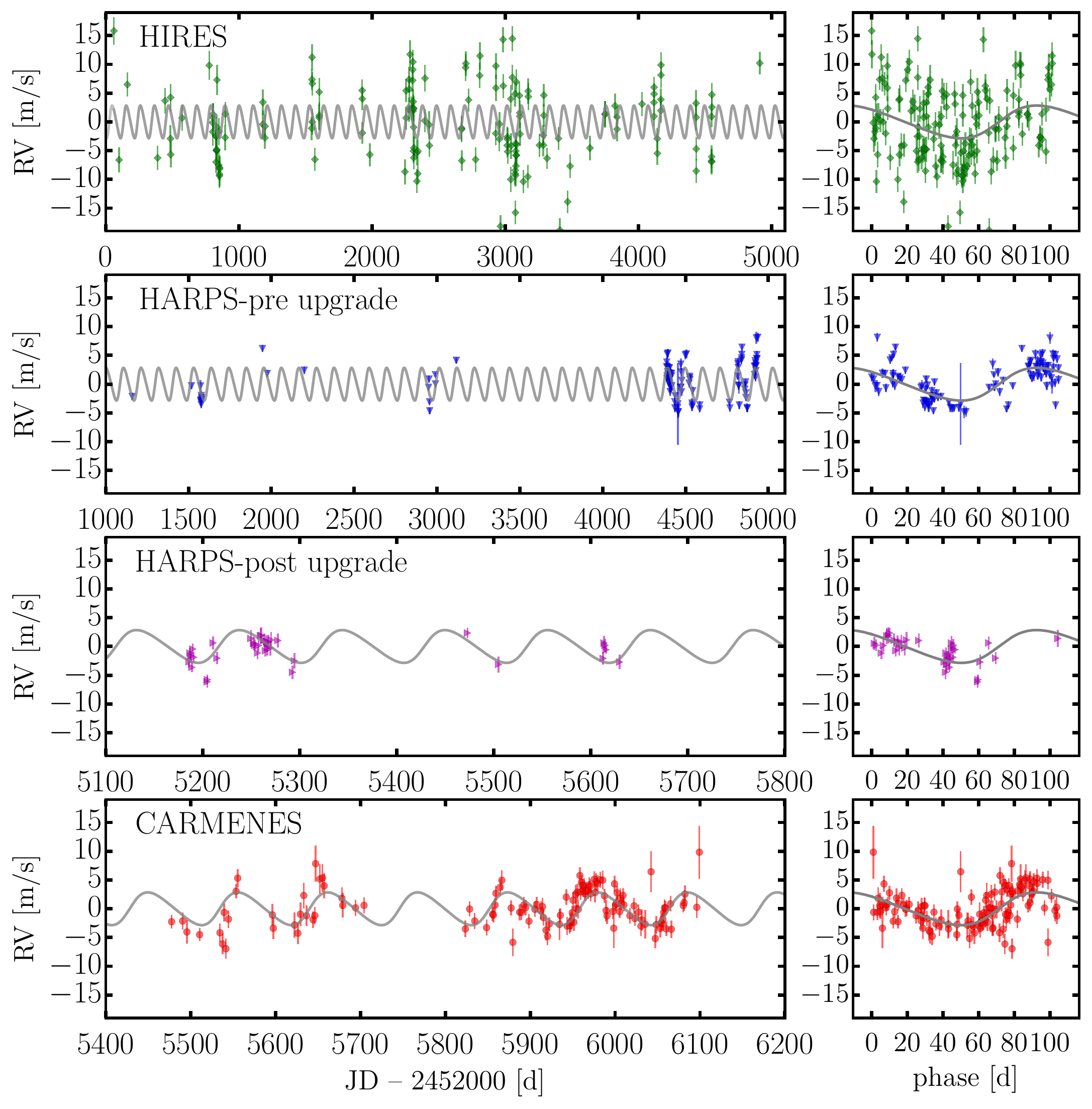}

\caption{Unfolded ({\em left}) and phase-folded ({\em right}) RV measurements of the individual instruments for HD\,180617, together with the best fit overplotted.}
\label{fig:kepfit_V1428}
\end{figure*}

\begin{figure}[]
\centering
\resizebox{\hsize}{!}{\includegraphics{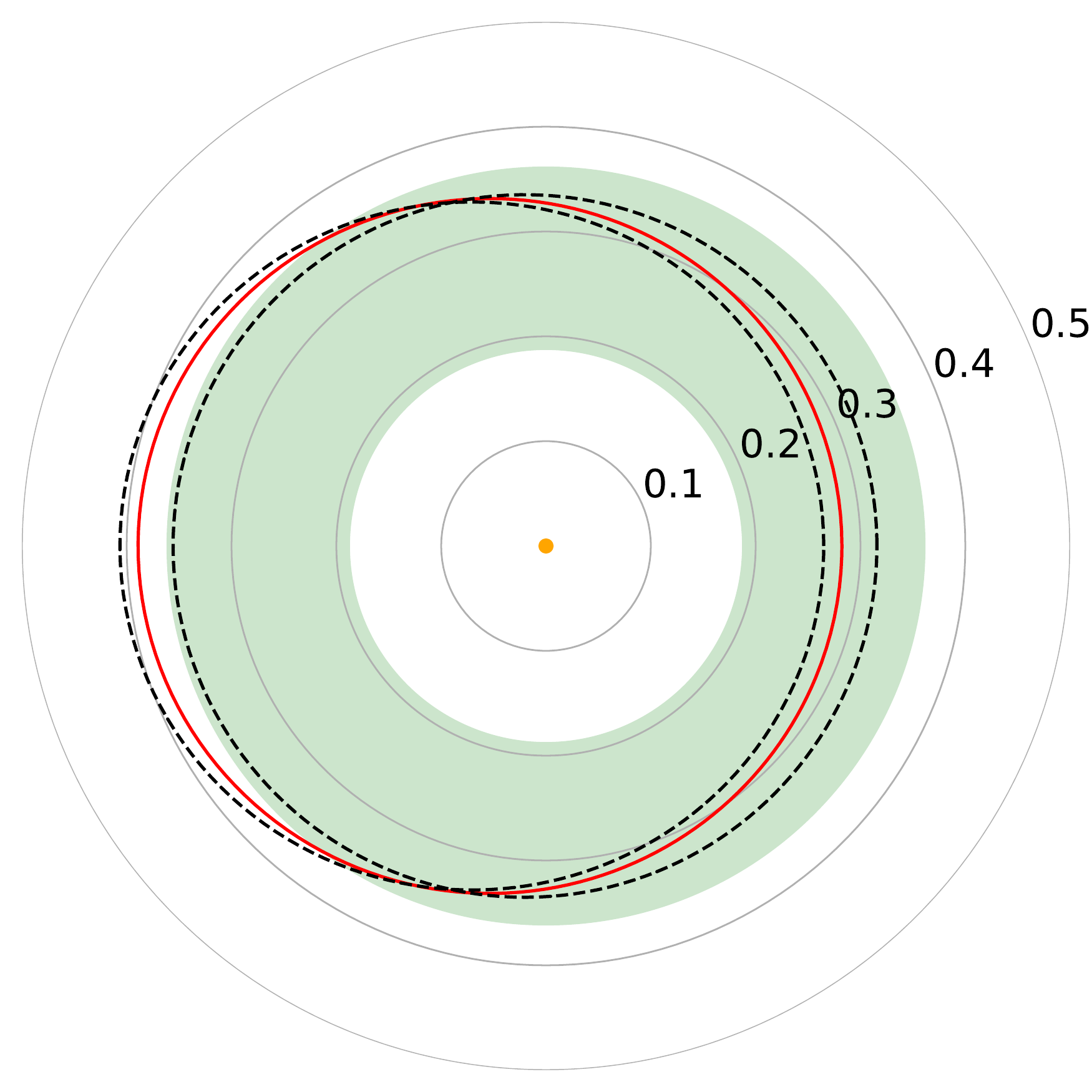}}

\caption{Orbital plot for HD\,180617 b around its host star. The red ellipse corresponds to the best-fit solution, while the dashed ellipses represent the orbits 
  at the upper and lower 1$\sigma$ uncertainty levels of the eccentricity. The conservative liquid water habitable zone according to \citet{Kopp_2014}, shaded in green, is given 
  by $0.19 \,\mathrm{au} \leq a \leq 0.36 \,\mathrm{au}$. Separations are in units of astronomical units.
}
\label{fig:orbitplots_both}
\end{figure}

For the subsequent analysis we applied a Keplerian model to the combined data set. It included the following free parameters: semi-amplitude $K$, orbital period $P$, eccentricity $e$,
mean anomaly $M$ (valid for JD=2452061.959, the first observational RV epoch),
longitude of periastron $\varpi$, and RV offsets $\gamma$ for each individual instrument. The HARPS instrument underwent an upgrade in May 2015, during
which new optical fibers were implemented \citep{LoCurto2015}. In order to take this into account, we divided the HARPS data into two subsets, and applied two different RV offsets for the 
pre- and post-upgrade epochs, respectively. Although HIRES also was upgraded in August 2004,
when it received a new CCD, we did not subdivide its data, as \citet{Butler_2017} showed that the upgrade did not produce a significant RV offset.\\ \indent
All free parameters were determined by minimizing the negative logarithm of the model likelihood, based on the Nelder-Mead simplex algorithm \citep{Nelder_1965}.
The resulting quantities, the time of transit $t_{\rm trans.}$, and the instrument-dependent RV jitter variances that were modeled simultaneously during the parameter optimization following \citet{Baluev_2009} are listed in Table~\ref{tab:kepfit_V1428}. 
The $1\sigma$ uncertainties were estimated from the posterior distributions derived by Markov chain Monte Carlo (MCMC) 
sampling (see Fig.~\ref{fig:fiterrors_V1428}). The same free parameters as in the Keplerian model, together with flat priors, were employed for the MCMC analysis, for which we made use of the open-source affine-invariant 
ensemble sampler \texttt{emcee} \citep{Foreman_2013}. Plots of the RV time series are shown in Fig.~\ref{fig:kepfit_V1428}.\\ \indent
\begin{table}[]
 \caption{Best-fit parameters with the corresponding 1$\sigma$ errors for HD\,180617 b.
 }  
\label{tab:kepfit_V1428}  
\centering      
\begin{tabular}{l  r r r r r r r r}     
\hline 
\noalign{\vskip 0.5mm}
\hline
\noalign{\vskip 0.5mm}

Orbital Parameters & \hspace{20 mm} HD\,180617 b  \\     
\hline 
\noalign{\vskip 0.9mm}

$K_b$ [m/s] & 2.85 $_{- 0.25}^{ + 0.16}$        \\  \noalign{\vskip 0.9mm}
$P_b$ [day] & 105.90 $_{- 0.10}^{ + 0.09}$        \\  \noalign{\vskip 0.9mm}
$e_b$ & 0.16 $_{- 0.10}^{ + 0.05}$        \\  \noalign{\vskip 0.9mm}
$\varpi_b$ [deg] & 269 $_{- 45}^{ + 30}$        \\  \noalign{\vskip 0.9mm}
$M_b$ [deg]$^{a}$ & 118 $_{- 37}^{ + 49}$        \\  \noalign{\vskip 0.9mm}
$l_b$ [deg]$^{b}$ & 27 $_{- 15}^{ + 14}$        \\  \noalign{\vskip 0.9mm}
$t_{\rm trans.}$ [BJD] & 2451974.5 $_{- 4.1}^{ + 4.5}$        \\  \noalign{\vskip 3.5mm}
$\gamma_{\rm HIRES}$ [m/s] & 0.43 $_{- 0.48}^{ + 0.47}$        \\  \noalign{\vskip 0.9mm}
$\gamma_{\rm HARPS-pre}$ [m/s] & -0.44 $_{- 0.21}^{ + 0.24}$        \\  \noalign{\vskip 0.9mm}
$\gamma_{\rm HARPS-post}$ [m/s] & -4.67 $_{- 0.37}^{ + 0.30}$        \\  \noalign{\vskip 0.9mm}
$\gamma_{\rm CARM.}$ [m/s] & -0.41 $_{- 0.22}^{ + 0.21}$        \\  \noalign{\vskip 0.9mm}
$\sigma_{\rm jitt,HIRES}$ [m/s] & 5.62 $_{- 0.29}^{ + 0.49}$        \\  \noalign{\vskip 0.9mm}
$\sigma_{\rm jitt,HARPS-pre}$ [m/s] & 2.03 $_{- 0.11}^{ + 0.21}$        \\  \noalign{\vskip 0.9mm}
$\sigma_{\rm jitt,HARPS-post}$ [m/s] & 1.41 $_{- 0.13}^{ + 0.49}$        \\  \noalign{\vskip 0.9mm}
$\sigma_{\rm jitt,CARM.}$ [m/s] & 1.69 $_{- 0.16}^{ + 0.27}$        \\  \noalign{\vskip 3.5mm}
$m_b \sin i$ [M$_\oplus$] & 12.2 $_{- 1.4}^{ + 1.0}$        \\  \noalign{\vskip 0.9mm}
$a_b$ [au] & 0.3357 $_{- 0.0100}^{ + 0.0099}$        \\  \noalign{\vskip 0.9mm}
\noalign{\vskip 0.5mm}
\hline 
\end{tabular} 
\begin{list}{}{}
\item[$^{a}$] The mean anomaly is valid for the first epoch of observation at JD=2452061.959.
\item[$^{b}$] The mean longitude is defined by $l_b = (\varpi_b+M_b)$ modulus 360\,deg.
\end{list}
\end{table}
In combination with the stellar parameters in Table~\ref{table.stars}, the measured semi-amplitude $K_b=2.85_{-0.25}^{+0.16}\,\metrepersecondnp$ and the period of $P_b = 105.90_{-0.10}^{+0.09}$~\,d translate into a planet with a minimum 
mass of $m_p \sin i = 12.2_{-1.4}^{+1.0}$\,~M$_\oplus$,
a semi-major axis of $a_b=0.3357_{-0.0100}^{+0.0099}$\,~au, and an eccentricity of  $e_b=0.16_{-0.10}^{+0.05}$. 
The errors of the mass and the semi-major axis take the uncertainty of the stellar mass into account.
The eccentricity is not well constrained and has a high uncertainty toward
lower values. This is illustrated in the posterior distrubutions from the MCMC analysis. The distribution of $e_b$ is concentrated at lower values with a 
mean of $e_b=0.14_{-0.08}^{+0.07}$.
According to \citet{Kopp_2014}, the best-fit solution places the planet at the outer edge of the conservative estimate of 
the liquid water habitable zone for its host star, $0.19 \,\mathrm{au} \leq a \leq 0.36 \,\mathrm{au}$. If the eccentricity is non-zero, the planet is inside the HZ for about
half of the orbit, near periastron, while it is outside for the other half, near apastron (see Fig.~\ref{fig:orbitplots_both}).
The stellar irradiance at the top of the planet atmosphere varies between 
$S_\mathrm{max}\approx 0.41$\,S$_0$ and S$_\mathrm{min}\approx 0.22$\,S$_0$, where the solar constant S$_0$ = 1360.8\,W\,m$^{-2}$ is the solar irradiance at the distance of 1 au.\\ \indent
For CARMENES and the HARPS data sets, the total RV jitter is comparable at $\lesssim2\,\metrepersecondnp$, while it is 5.62\,$\metrepersecondnp$ for the HIRES data.
With 1.77\,$\metrepersecondnp$ , the root mean square of the residuals of the model fits is lowest for the HARPS post-upgrade data, followed by the HARPS pre-upgrade (2.15\,$\metrepersecondnp$) and the CARMENES (2.66\,$\metrepersecondnp$) data. Again, HIRES shows the 
highest value, with 6.02\,$\metrepersecondnp$. Compared to the formal uncertainties of the measurements (see Sect.~\ref{sec:data}), this means a factor of around three for HIRES and the HARPS pre-upgrade, and a factor of two for the HARPS post-upgrade and the CARMENES data.\\ \indent
The GLS periodogram of the residuals does not show any significant signals (lower panel Fig.~\ref{fig:gls_RV_V1428}).
However, there is some power at periods around 45 days and its overtones. This can be attributed to the 
rotation period $P_{\rm rot} = 46.04$\,d of the host star. Moreover, the strongest peak at around $P\approx 190$\,d nearly reaches the 1\% FAP level. The source of this signal is currently
unknown, although the activity indicators hint at stellar activity.
We also analyzed the residuals for possible long-term structures, but found the results inconclusive. While the entire data set shows only a negligible drift term of 0.03\,ms$^{-1}$yr$^{-1}$, it is more significant for the residuals of the most recent RV measurements.
A linear fit to the combined residuals from the HARPS post-upgrade epoch together with the CARMENES data gives a slope of around 0.5\,ms$^{-1}$yr$^{-1}$. This could be an indication for an additional long-term signal, which in the HIRES data might be obscured by the comparably 
high scatter.\\ \indent
In order to investigate the robustness of our fit results against possible contributions from correlated noise 
and additional RV or activity signals, we studied five additional models, using the concepts of moving averages (MA) and Gaussian process (GP) regression. The results from this analysis and their discussion can be found in Appendix.~\ref{app:Models}. 
They agree with the orbital parameters reported here.
\begin{figure}[]
\resizebox{\hsize}{!}{\includegraphics[]{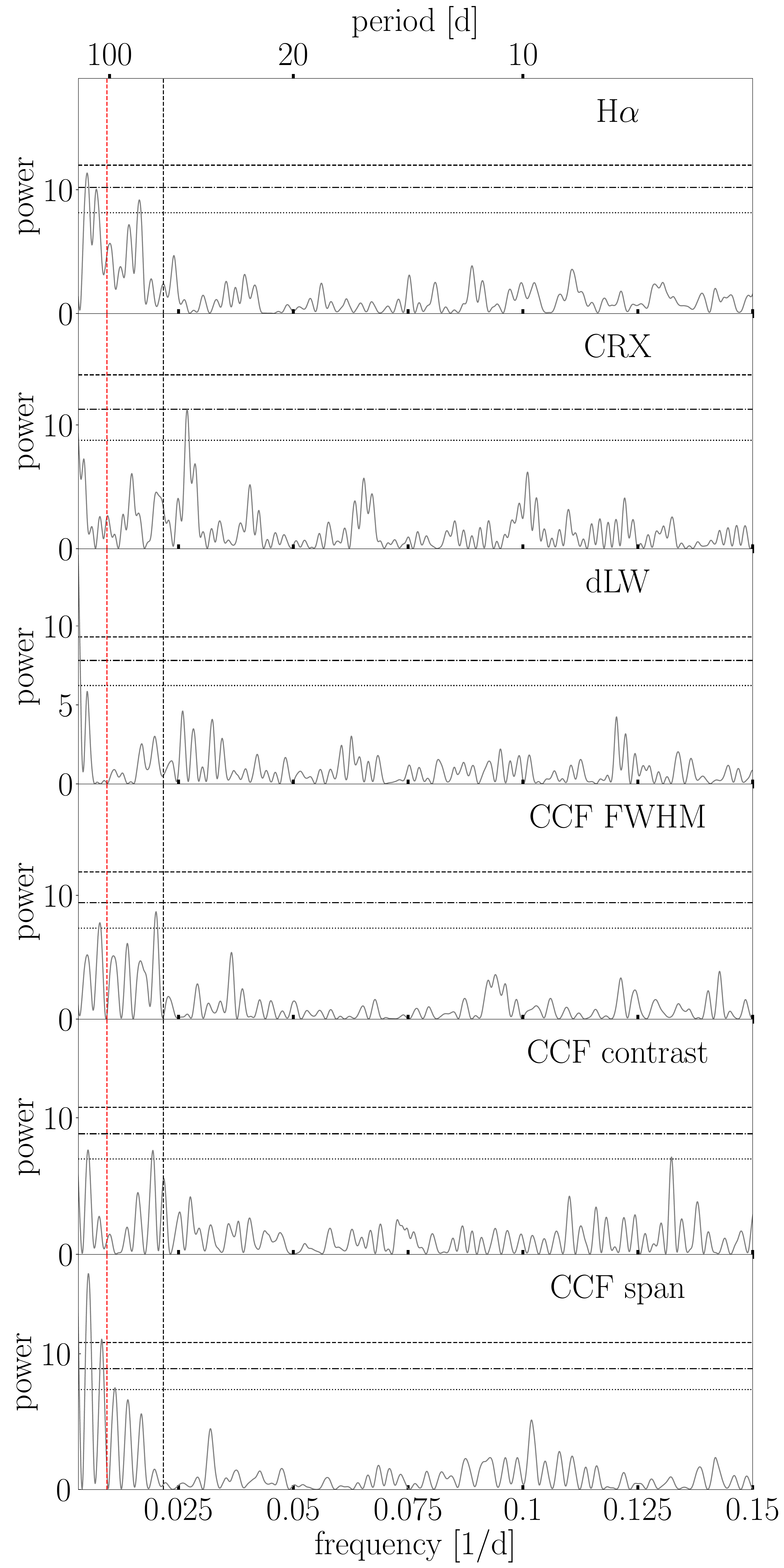}}

\caption{GLS periodograms of activity indicators derived from the CARMENES spectra of HD\,180617 with lines (dotted, dash-dotted, and dashed) at FAP levels of 10\%, 1\%, and 0.1\%. The vertical red and black lines represent the best-fit planet period 
at $P_b = $ 105.9\,d, and the rotation period of the star at $P_\mathrm{rot} = $ 46.04\,d, respectively. No significant power is found at the planetary period. 
}
\label{fig:gls_activity_V1428}
\end{figure}  
\subsection{Activity analysis}\label{ssec:RV_activity}
Periodic variability in RV measurements can also be introduced by stellar activity. Chromospheric emission, chromatic dependence of the
RVs, and changes in line profiles are indicative of stellar activity. In Fig.~\ref{fig:gls_activity_V1428} we provide GLS periodograms for some of these quantities, derived from the available CARMENES spectra. The H$\alpha$ index, the 
chromatic index (CRX), and the differential line width (dLW), shown in the three uppermost panels, are derived from the time series of activity indicators directly provided by SERVAL \citep{Zechmeister_2017}. 
While the chromatic index reflects 
wavelength dependencies of the measured RVs in the different
\'echelle orders, which can result from the temperature surface structure on the star (spots or faculae), the differential line width serves as another measure for variations in the shape of the line profiles.
In addition, we evaluated the full width at half-maximum (FWHM) of the cross-correlation function (CCF). We show periodograms of the CCF-FWHM, the contrast, and the bisector span. The latter is defined 
as the difference between the average bisector values in the two CCF regions from 90\% to 60\% and from 40\% to 10\%.\\ \indent
There is no significant power at any of these indicators at the orbital period $P_b = $ 105.90\,d.
The highest peak in the GLS periodogram of H$\alpha$ is at a period of $P\approx 196$\,d ($f\approx 0.00510$\,d$^{-1}$) and an FAP of nearly 0.1\%. 
This period is consistent with a peak found in the periodogram of the RV residuals after subtracting the best-fit planet solution.
The second highest peak at around 140\,d ($f\approx 0.00716$\,d$^{-1}$) and an FAP around 1\% is an alias. It disappears after subtracting a fit to the 196\,d period.
The sampling function (not plotted) shows a strong signal at around 436 days ($f_s\approx 0.00229$\,d$^{-1}$). Together with that sampling frequency $f_s$, the signal at $f\approx 0.00510$\,d$^{-1}$ could produce an 
alias at around $f_a = f+f_s$. The indicators sensitive to line profile variations also show some power at around 200 days.
Peaks (FAP$\sim$10\%) at around 50\,d, which is close to the rotation period of the star, are evident in the periodograms of CCF-FWHM and CCF contrast.\\ \indent
We also investigated the activity indicators from the other instruments. For HARPS we analyzed the time series of the H$\alpha$ and CRX indices, the dLW, the CCF-FWHM, and the bisector span.
The HIRES data provided time series of the S-index and the H-index, two further chromospheric activity indicators \citep{Gomes_2011, Butler_2017}. Signatures
from the stellar rotation are obvious in HARPS H$\alpha$ data with an FAP below 0.1\% around the period of 50\,d and recognizable although insignificant (FAP$\sim$10\%) in the CCF-FWHM periodogram. The same is true for the HIRES indices. 
In the GLS of the S-index the rotational period of the star was recovered from a signal with an FAP below 1\%, while a forest of peaks with low significance (FAP$\gtrsim$10\%) was found in the H-index data.
Furthermore, the periodograms of the H-index and the dLW from HARPS also showed some power at periods between 185\,d and 195\,d and some significant
power excess was found in the GLS of the S-index at around 3500\,d, which might imply a long-term magnetic cycle.
However, none of these additional indicators showed any significant signal at the planetary period, and we conclude that the RV signal at 105.90\,d is due to the stellar reflex motion caused by a planetary companion rather 
than stellar activity.
We found no correlation between any of the activity indicators and the RV model residuals either.

\section{Summary and concluding remarks}\label{sec:conclusions}
We presented radial velocity time series of the inactive early-M dwarf HD\,180617 using RV measurements from the
visual channel of the high-resolution spectrograph CARMENES in addition to HARPS and HIRES archival data. The combined data sets indicate the existence of a planet with a minimum mass of 12.2\,M$_\oplus$ on
a 105.9\,d orbit around its stellar host. Our investigation of periodicities in the activity indicators for H$\alpha$ emission, RV chromaticity, and line profile variations shows no obvious indication that the planetary signal 
is due to activity-induced RV variability. The orbit of the Neptune-like planet, 
with semi-major axis $a=0.3357$\,~au and eccentricity $e = 0.16$, is partly located in its host star's liquid water habitable zone.
Under the condition of an edge-on view onto the system, the semi-amplitude astrometric signature at Earth's distance is estimated to be about 4.6$\,\mu$as.
With decreasing orbital inclination, the planetary mass and therefore the astrometric signature increase, therefore this is only a lower limit and the astrometric orbit could potentially be observed by \textit{Gaia}. 
Because of the proximity of HD\,180617 ($d=5.9$\,pc), the angular separation of the planet from its host star is rather large, comparable to the ${\sim}50$\,mas inner working angle of a high-performance coronograph on a 4~m class space telescope. 
While the mass of HD\,180617\,b is too high to be considered a close Earth analog, it is thus nevertheless a prime target for future missions - such as the HabEx concept \citep{Menn_2016} - aimed at studying the atmospheres of potentially life-bearing planets, 
or ESO's ground-based Extremely Large Telescope \citep{Gil_2007}.\\ \indent
Only 15 closer stars, 10 of which are M dwarfs, are currently known to have an exoplanet. Within this group, HD\,180617 is the seventh brightest in the $V$ band.
Moreover, it is the primary of the wide binary containing the low-mass star vB\,10. We provide new measurements on the binary separation and new determinations of the primary mass, radius, luminosity, and Galactocentric space motion.
Of the 88 currently confirmed binary systems that host exoplanets, only a handful possess an M dwarf as the primary \citep{Schwarz_2016}. Except for Gliese\,676\,A/B and Gliese\,15\,A/B, the system investigated in this work is the third of this kind with an exoplanet
in an S-type\footnote{In this context, ``S'' stands for satellite.} configuration \citep{Rabl_1988}, which means that the planet is in a close orbit around only one of the systems' stars. Because of the wide separation ($s \ge$ 450\,au) between the system's stellar components (more than 1300 times larger than the semi-major axis of HD\,180617\,b), 
the perturbation of vB\,10 on the formation and dynamical evolution of the planet is expected to be negligible \citep[e.g.,][]{Whit_1998, Quin_2007, Quin_2008}. In this configuration, the gravitational pull from the host star on the planet is about 10$^7$ times stronger than from the secondary. Furthermore, given the
very low luminosity of 425$\pm$4~10$^{-6}$\,L$_\odot$ of the M8.0\,V companion \citep{Cif_2017}, the presence of vB\,10 is not expected to affect the radiative equilibrium of HD\,180617\,b. Consequently, in this particular case, binarity does not affect habitability.\\ \indent
In line with HD\,147379\,b, the first exoplanet discovery by CARMENES \citep{Reiners_2018a}, our work demonstrates the potential and capability of the instrument of finding exoplanets within the mass range of Neptunes and mini-Neptunes
within the habitable zones of M dwarfs. Both HD\,180617\,b and HD\,147379\,b have orbital periods of about 100\,d. Although dozens of exoplanets with lower masses on shorter orbits around M dwarfs are known, these two discoveries belong to
only a few known Neptunes at longer orbits with periods $\gtrsim 100$\,d. While they improve the sampling within this barely filled region of the exoplanet parameter space, even longer periods for exoplanets of 
similar masses around M dwarfs are becoming accessible as a result of the steadily increasing time baselines and the improved precision of current and future instruments.

\begin{acknowledgements}
CARMENES is an instrument for the Centro Astron\'omico Hispano-Alem\'an de
Calar Alto (CAHA, Almer\'{\i}a, Spain). 
CARMENES is funded by the German Max-Planck-Gesellschaft (MPG), 
the Spanish Consejo Superior de Investigaciones Cient\'{\i}ficas (CSIC),
the European Union through FEDER/ERF FICTS-2011-02 funds, 
and the members of the CARMENES Consortium 
(Max-Planck-Institut f\"ur Astronomie,
Instituto de Astrof\'{\i}sica de Andaluc\'{\i}a,
Landessternwarte K\"onigstuhl,
Institut de Ci\`encies de l'Espai,
Insitut f\"ur Astrophysik G\"ottingen,
Universidad Complutense de Madrid,
Th\"uringer Landessternwarte Tautenburg,
Instituto de Astrof\'{\i}sica de Canarias,
Hamburger Sternwarte,
Centro de Astrobiolog\'{\i}a and
Centro Astron\'omico Hispano-Alem\'an), 
with additional contributions by the Spanish Ministry of Science through projects AYA2016-79425-C3-1/2/3-P, ESP2016-80435-C2-1-R, and AYA2015-69350-C3-2-P,
the German Science Foundation through the Major Research Instrumentation 
  Programme and DFG Research Unit FOR2544 ``Blue Planets around Red Stars'', 
the Klaus Tschira Stiftung, 
the states of Baden-W\"urttemberg and Niedersachsen, 
and by the Junta de Andaluc\'{\i}a.
Based on data from the CARMENES data archive at CAB (INTA-CSIC).
N.\,A.\,D. acknowledges support from FONDECYT \#3180063.
Based on observations collected at the European Organisation for Astronomical Research in the Southern Hemisphere under ESO programme(s) 072.C-0488(E), 183.C-0437(A), 191.C-0873(A), 191.C-0873(B), 191.C-0873(D),
191.C-0873(E), 191.C-0873(F).
Some of the data presented herein were obtained at the W. M. Keck Observatory, which is operated as a scientific partnership among the California Institute of Technology, 
the University of California and the National Aeronautics and Space Administration. The Observatory was made possible by the generous financial support of the W. M. Keck Foundation.
We made use of the \texttt{corner.py} package \citep{Foreman_2016}.
We thank the anonymous referee for useful comments that improved the paper.
\end{acknowledgements}

\bibliographystyle{aa}
\bibliography{ak_V1428}

\begin{appendix} 

\clearpage
\section{MCMC analysis}
\begin{figure*}[]
\centering
\includegraphics[width=17cm]{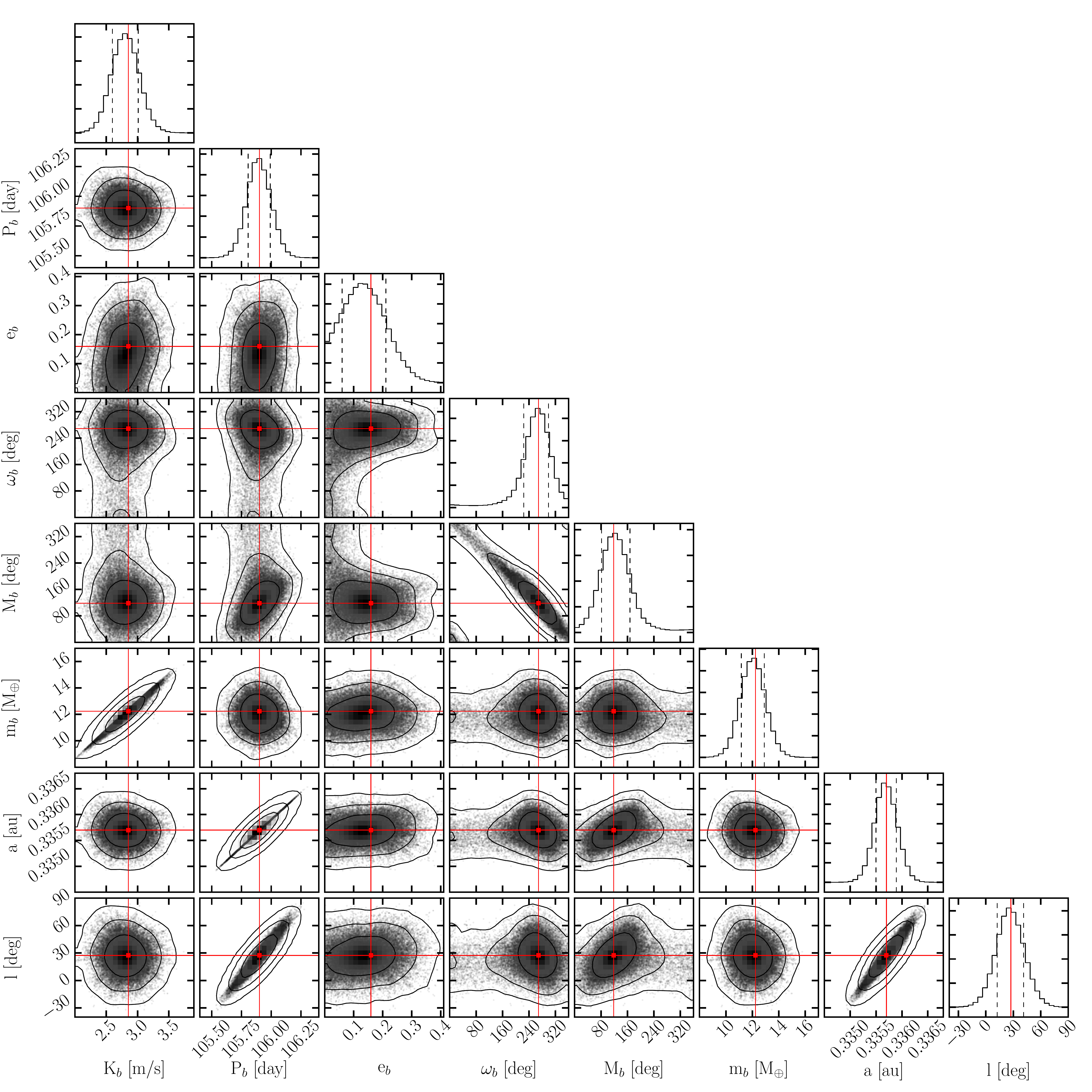}
\caption{
Posterior distributions from the MCMC analysis on HD\,180617 b. The best-fit solution is indicated by red lines and the contours define uncertainty levels of 1, 2, and 3$\sigma$ with respect to the mean.
}
\label{fig:fiterrors_V1428} 
\end{figure*}

\section{Model comparison}\label{app:Models}
In order to investigate the influence of the spurious RV and activity signals on the estimation of the orbital parameters, we applied five additional models to the full data set for comparison.
First, a moving-average (MA) noise model \citep{Tuo_2013a, Tuo_2013b} was used to assess the effect of correlated noise. By the means of two additional parameters, an amplitude and a correlation length, this method accounts for the influence of consecutive data 
points on each another.
In our second approach, the possible contributions were modeled by Gaussian process (GP) regressions, which are now widely used within the community \citep[e.g.,][]{Raj_2015, Dum_2017}, and for which we used two different packages. The first was the python 
library {\tt george}\footnote{\tt http://dfm.io/george} developed by D. Foreman-Mackey \citep{Ambi_2015}. Here the covariance matrix was calculated over a quasi-periodic kernel, which consists of four hyperparameters.
They are commonly translated into an RV amplitude $h$, a rotational period $\Theta$, a weighting factor $w$, and a timescale $\lambda$, which can be associated with the corresponding signal's lifetime, or in particular cases with the lifetime of its underlying process, like
an active region within the stellar photosphere \citep[e.g.,][]{Per_2017}. An MCMC analysis was applied to estimate the uncertainties.
The second package was the fast and scalable GP regression package {\tt celerite}\footnote{\tt https://github.com/dfm/celerite}, which makes use of a damped harmonic oscillator as its kernel \citep{Foreman_2017}. This GP model is characterized by the natural oscillator frequency, 
 or period $P_0$, 
damping time $\tau$, and peak height $S_0$, which scales with the power at the corresponding frequency. The GP model can either be constrained by the RV data alone or by supplementary data, like in our case the time series of the activity indicator H$\alpha$ 
from the CARMENES data.\\ \indent
Within all of the models we accounted for the RV offsets between the instruments and for RV jitter terms. The likelihood $\ln L$ was calculated as an indicator for the significance of the different approaches.
The results from all the different methods are summarized in Table~\ref{table:models}. 
Model 1 lists the results from our original Keplerian fit (see Sect.~\ref{ssec:RV_model}). Model 2 with MA accounts for correlated noise, as well as for the planet via an additional Keplerian term. Within their errors, the determined orbital parameters agree with the 
parameters from model 1. Models 3 and 4, using {\tt george}, and models 5 and 6, using {\tt celerite}, are based on GP regressions. In model 3, only the correlated noise is modeled, and the period of the dominant RV signal ($P_b = 105.9$\,d) is fairly recovered.
The long timescale $\lambda$ = 502.54\,d indicates that the RV variation is stable over the time of the observations, as expected from a planet. High attention should be given to models 4--6, which all account for the planet and for correlated noise. 
Like model 2, they all agree with our original results from model 1. The parameter showing the highest dispersion is the eccentricity, which is quasi-zero in model 6, which was additionally constrained by the H$\alpha$ data. 
However, it is also the parmameter with the highest relative errors, and it has been discussed before that it is not well constrained in model 1, with a high uncertainty in particular toward low values. 
Within the hyperparameters of the GP models 4 and 5, the rotational period of the star is recovered.\\ \indent
Altogether, from the comparison of the different approaches, we conclude that the reported results for the orbital parameters in Sect.~\ref{ssec:RV_model} are robust against and not significantly affected by the additional signals in the RV time series.
Therefore, although the $\ln L$ is increased for the models that take correlated noise into account, we consider the results from the traditional approach sufficiently reliable.
\begin{table*}
\caption{Results from different models to the combined RV time series$^{a}$.}
         \label{table:models}
\centering
         \begin{tabular}{lc c c cc c cc}
\hline 
\hline
\noalign{\smallskip}
Method        & max $\ln L$   & MA   &\phantom{a}     & \multicolumn{2}{c}{GP ({\tt george})} &\phantom{a}& \multicolumn{2}{c}{GP ({\tt celerite})}   \\  \cmidrule{5-6} \cmidrule{8-9} \noalign{\vskip 0.9mm}
Model             & \#1          & \#2                         &&  \#3          & \#4                      && \#5      & \#6       \\
                  & Kepler       & Kepler + CN                 && CN & Kepler + CN                && Kepler + CN &    Kepler + CN + H$\alpha$\\ \noalign{\vskip 0.9mm}
            \hline 
            \noalign{\vskip 3.5mm}
$\ln L$             & $-$1088.4    & $-$1054.7    && $-$1102.3 & $-$1069.2 && $-$1068.8 & $-$1071.6  \\ \noalign{\vskip 3.5mm} 

$K_b$ [ms$^{-1}$]    & 2.85 $_{- 0.25}^{ + 0.16}$    & 2.78 $_{- 0.29}^{ + 0.29}$      && \dots & 2.96$^{+0.32}_{-0.33}$   && 2.90 $_{- 0.35}^{ + 0.35}$ & 2.91 $_{- 0.33}^{ + 0.33}$ \\ \noalign{\vskip 0.9mm}
$P_b$ [d]            & 105.90 $_{- 0.10}^{ + 0.09}$    & 105.95 $_{- 0.13}^{ + 0.13}$  && \dots & 105.84$^{+0.17}_{-0.16}$ && 105.96 $_{- 0.15}^{ + 0.15}$ & 105.99 $_{- 0.13}^{ + 0.13}$  \\ \noalign{\vskip 0.9mm}
$e_b$                & 0.16 $_{- 0.10}^{ + 0.05}$    & 0.13 $_{- 0.05}^{ + 0.10}$      && \dots & 0.26 $_{- 0.15}^{ + 0.18}$ && 0.07 $_{- 0.05}^{ + 0.10}$ & 0.01 $_{- 0.05}^{ + 0.09}$  \\ \noalign{\vskip 3.5mm}

$P_0$ [d]            & \dots    & \dots    && \dots & \dots && 45.8$^{+10.2}_{-14.5}$ & 62.5$^{+6.3}_{-6.3}$\\ \noalign{\vskip 0.9mm}
$\tau$ [d]           & \dots    & \dots    && \dots & \dots && 14.9$^{+22.8}_{-7.7}$ & 12.1$^{+1.5}_{-1.5}$ \\ \noalign{\vskip 0.9mm}
$S_0$     & \dots    & \dots    && \dots & \dots &&  17$_{- 9}^{ + 20}$ & 12 $_{- 2}^{ + 2}$ \\ \noalign{\vskip 3.5mm}

$\Theta$ [d]         & \dots    & \dots    && 104.7$^{+3.8}_{-1.9}$ & 49.0$^{+11.0}_{-3.9}$ &&  \dots & \dots \\ \noalign{\vskip 0.9mm}
$\lambda$ [d]        & \dots    & \dots    && 502.5$^{+271.4}_{-170.2}$ & 69.1 $^{+30.3}_{-14.7}$ && \dots     & \dots  \\ \noalign{\vskip 0.9mm}
$h$ [ms$^{-1}$]      & \dots    & \dots    && 2.3$^{+0.4}_{-0.3}$ & 0.8$^{+0.4}_{-0.4}$ &&  \dots & \dots \\ \noalign{\vskip 0.9mm}
$w$ [ms$^{-1}$]      & \dots    & \dots    && 1.6$^{+1.0}_{-0.6}$ & 2.3$^{+1.3}_{-0.7}$ &&  \dots & \dots \\ \noalign{\vskip 3.5mm}
        \noalign{\smallskip}
            \hline
\end{tabular}
\begin{list}{}{}
\item[$^{a}$] GP: Gaussian Process regression; MA: Moving Average; CN: Correlated Noise; H$\alpha$: H$\alpha$ constrained.
\end{list}
\end{table*}  
   
\section{RV data}

\begin{table}
\caption{Doppler measurements for HD\,180617. All RVs are corrected for barycentric motion and secular acceleration.} 
\label{table:V1428_RVs_a} 

\centering  

\begin{tabular}{c c c l } 

\hline\hline    
\noalign{\vskip 0.5mm}

JD & RV [m\,s$^{-1}$] & $\sigma_{RV}$ [m\,s$^{-1}$]   & Instrument \\  

\hline     
\noalign{\vskip 0.5mm}    

2452061.959   &   16.26   &   2.40   &   HIRES \\
2452100.972   &   -6.19   &   2.21   &   HIRES \\
2452162.856   &   6.93   &   2.07   &   HIRES \\
2452390.116   &   -5.83   &   2.10   &   HIRES \\
2452445.953   &   4.11   &   2.24   &   HIRES \\
2452486.838   &   -2.49   &   2.32   &   HIRES \\
2452487.825   &   4.69   &   2.37   &   HIRES \\
2452488.910   &   -5.26   &   2.24   &   HIRES \\
2452573.730   &   1.11   &   2.24   &   HIRES \\
2452778.075   &   10.24   &   2.55   &   HIRES \\
2452804.100   &   0.29   &   2.37   &   HIRES \\
2452805.022   &   -1.28   &   2.45   &   HIRES \\
2452806.068   &   1.20   &   2.40   &   HIRES \\
2452806.964   &   1.76   &   1.95   &   HIRES \\
2452828.859   &   -2.10   &   2.19   &   HIRES \\
2452829.068   &   -6.54   &   1.95   &   HIRES \\
2452830.078   &   -2.07   &   2.21   &   HIRES \\
2452831.994   &   -4.58   &   2.13   &   HIRES \\
2452832.832   &   -4.51   &   2.01   &   HIRES \\
2452833.942   &   -4.36   &   2.19   &   HIRES \\
2452835.026   &   -0.50   &   2.16   &   HIRES \\
2452836.025   &   7.73   &   2.59   &   HIRES \\
2452848.916   &   -7.07   &   2.50   &   HIRES \\
2452851.006   &   -1.25   &   2.35   &   HIRES \\
2452853.998   &   -8.94   &   2.19   &   HIRES \\
2452854.976   &   -7.51   &   2.16   &   HIRES \\
2452854.985   &   -8.66   &   2.16   &   HIRES \\
2452855.976   &   -7.74   &   2.10   &   HIRES \\
2452897.744   &   -2.30   &   2.19   &   HIRES \\
2452898.829   &   1.80   &   2.04   &   HIRES \\
2453159.811   &   -2.59   &   0.64   &   HARPS-pre \\
2453179.862   &   -0.00   &   2.40   &   HIRES \\
2453181.044   &   3.83   &   2.27   &   HIRES \\
2453195.860   &   -2.34   &   1.92   &   HIRES \\
2453196.858   &   -0.34   &   2.04   &   HIRES \\
2453517.840   &   -0.74   &   0.58   &   HARPS-pre \\
2453546.939   &   -0.69   &   2.07   &   HIRES \\
2453549.003   &   7.79   &   1.83   &   HIRES \\
2453549.926   &   11.66   &   2.32   &   HIRES \\
2453550.991   &   0.42   &   2.21   &   HIRES \\
2453551.928   &   7.09   &   2.07   &   HIRES \\
2453570.988   &   -6.10   &   2.01   &   HIRES \\
2453572.766   &   -0.76   &   1.06   &   HARPS-pre \\
2453573.723   &   -3.23   &   0.54   &   HARPS-pre \\
2453574.686   &   -3.33   &   0.61   &   HARPS-pre \\
2453575.651   &   -3.57   &   0.58   &   HARPS-pre \\
2453576.654   &   -3.99   &   0.64   &   HARPS-pre \\
2453577.679   &   -2.41   &   0.58   &   HARPS-pre \\
2453578.676   &   -2.80   &   0.61   &   HARPS-pre \\
2453579.658   &   -3.08   &   0.50   &   HARPS-pre \\
2453602.889   &   1.19   &   2.24   &   HIRES \\
2453603.915   &   5.63   &   2.42   &   HIRES \\
2453604.792   &   1.53   &   2.35   &   HIRES \\
2453926.056   &   5.90   &   1.89   &   HIRES \\
2453926.982   &   4.44   &   1.98   &   HIRES \\
2453927.921   &   -1.49   &   1.92   &   HIRES \\
2453946.578   &   5.76   &   0.61   &   HARPS-pre \\
2453975.589   &   1.41   &   0.58   &   HARPS-pre \\
2453983.766   &   -5.27   &   2.01   &   HIRES \\
2454198.906   &   1.99   &   0.68   &   HARPS-pre \\
  
\hline           
\end{tabular}
\end{table}

\begin{table}
\caption*{Doppler measurements for HD\,180617 -- continued.} 
\vspace{\baselineskip}
\label{table:V1428_RVs_b} 

\centering  

\begin{tabular}{c c c l } 

\hline\hline    
\noalign{\vskip 0.5mm}

JD & RV [m\,s$^{-1}$] & $\sigma_{RV}$ [m\,s$^{-1}$]   & Instrument \\   

\hline     
\noalign{\vskip 0.5mm}    

2454247.059   &   -8.24   &   2.19   &   HIRES \\
2454255.945   &   1.12   &   2.01   &   HIRES \\
2454255.950   &   5.86   &   2.01   &   HIRES \\
2454277.872   &   2.57   &   2.07   &   HIRES \\
2454278.925   &   7.79   &   2.19   &   HIRES \\
2454279.958   &   6.11   &   2.13   &   HIRES \\
2454286.023   &   12.17   &   2.47   &   HIRES \\
2454294.920   &   2.77   &   2.07   &   HIRES \\
2454304.990   &   7.64   &   1.98   &   HIRES \\
2454305.996   &   9.50   &   2.07   &   HIRES \\
2454306.963   &   10.86   &   2.10   &   HIRES \\
2454308.032   &   2.33   &   2.32   &   HIRES \\
2454308.989   &   3.94   &   2.16   &   HIRES \\
2454309.990   &   2.83   &   2.35   &   HIRES \\
2454310.980   &   -4.64   &   2.07   &   HIRES \\
2454311.975   &   -0.50   &   2.13   &   HIRES \\
2454312.971   &   2.71   &   2.21   &   HIRES \\
2454313.967   &   -5.80   &   2.07   &   HIRES \\
2454315.016   &   -4.10   &   2.01   &   HIRES \\
2454335.960   &   -4.94   &   2.24   &   HIRES \\
2454336.994   &   -9.88   &   2.10   &   HIRES \\
2454343.892   &   -8.58   &   1.83   &   HIRES \\
2454344.955   &   -4.66   &   2.10   &   HIRES \\
2454396.725   &   0.63   &   2.01   &   HIRES \\
2454397.751   &   8.01   &   2.32   &   HIRES \\
2454429.714   &   -0.07   &   2.40   &   HIRES \\
2454430.717   &   -3.68   &   1.95   &   HIRES \\
2454673.083   &   -0.66   &   2.16   &   HIRES \\
2454673.091   &   -0.87   &   2.24   &   HIRES \\
2454673.957   &   -6.31   &   1.83   &   HIRES \\
2454702.900   &   9.88   &   2.37   &   HIRES \\
2454704.937   &   10.58   &   2.13   &   HIRES \\
2454724.810   &   4.25   &   2.01   &   HIRES \\
2454777.766   &   -6.11   &   2.13   &   HIRES \\
2454808.701   &   8.48   &   2.19   &   HIRES \\
2454810.711   &   11.91   &   2.40   &   HIRES \\
2454929.127   &   -1.58   &   2.07   &   HIRES \\
2454930.073   &   10.14   &   2.27   &   HIRES \\
2454930.079   &   6.33   &   2.24   &   HIRES \\
2454948.911   &   -1.01   &   0.54   &   HARPS-pre \\
2454951.912   &   0.44   &   0.58   &   HARPS-pre \\
2454954.905   &   -3.52   &   0.50   &   HARPS-pre \\
2454955.884   &   -5.08   &   0.58   &   HARPS-pre \\
2454956.108   &   -2.55   &   2.13   &   HIRES \\
2454957.019   &   0.05   &   2.04   &   HIRES \\
2454964.113   &   -17.73   &   1.98   &   HIRES \\
2454983.952   &   14.74   &   2.19   &   HIRES \\
2454985.098   &   6.69   &   2.13   &   HIRES \\
2454989.868   &   1.22   &   0.40   &   HARPS-pre \\
2454990.853   &   -0.40   &   0.61   &   HARPS-pre \\
2455015.005   &   -3.66   &   2.07   &   HIRES \\
2455015.886   &   -2.19   &   1.80   &   HIRES \\
2455016.960   &   -2.57   &   2.01   &   HIRES \\
2455019.066   &   -4.74   &   2.16   &   HIRES \\
2455041.869   &   -9.58   &   2.24   &   HIRES \\
2455042.916   &   4.43   &   2.40   &   HIRES \\
2455043.862   &   -3.74   &   2.24   &   HIRES \\
2455053.027   &   8.16   &   2.21   &   HIRES \\
2455053.034   &   14.91   &   2.19   &   HIRES \\
2455073.742   &   5.27   &   2.21   &   HIRES \\
  
\hline           
\end{tabular}
\end{table}

\clearpage

\begin{table}
\caption*{Doppler measurements for HD\,180617 -- continued.}
\label{table:V1428_RVs_c} 

\centering  

\begin{tabular}{c c c l } 

\hline\hline    
\noalign{\vskip 0.5mm}

JD & RV [m\,s$^{-1}$] & $\sigma_{RV}$ [m\,s$^{-1}$]   & Instrument \\  

\hline     
\noalign{\vskip 0.5mm}    

2455074.740   &   -8.58   &   2.04   &   HIRES \\
2455075.729   &   -7.98   &   2.42   &   HIRES \\
2455076.736   &   -15.33   &   2.07   &   HIRES \\
2455077.733   &   -8.90   &   1.95   &   HIRES \\
2455078.735   &   -4.65   &   2.07   &   HIRES \\
2455079.720   &   -3.32   &   1.92   &   HIRES \\
2455080.732   &   -7.25   &   1.92   &   HIRES \\
2455081.727   &   -5.38   &   1.98   &   HIRES \\
2455082.720   &   7.40   &   2.16   &   HIRES \\
2455083.726   &   -3.53   &   1.92   &   HIRES \\
2455084.722   &   -5.36   &   1.95   &   HIRES \\
2455106.753   &   5.01   &   2.29   &   HIRES \\
2455109.729   &   2.01   &   2.21   &   HIRES \\
2455109.735   &   1.40   &   2.04   &   HIRES \\
2455111.714   &   -1.68   &   2.01   &   HIRES \\
2455116.514   &   3.69   &   0.50   &   HARPS-pre \\
2455135.733   &   -9.94   &   1.98   &   HIRES \\
2455169.695   &   -9.10   &   2.01   &   HIRES \\
2455172.701   &   5.78   &   2.24   &   HIRES \\
2455173.699   &   4.83   &   2.35   &   HIRES \\
2455257.164   &   -3.62   &   2.01   &   HIRES \\
2455286.139   &   -1.92   &   1.80   &   HIRES \\
2455286.144   &   -3.62   &   1.95   &   HIRES \\
2455290.112   &   5.05   &   1.98   &   HIRES \\
2455290.117   &   1.57   &   1.98   &   HIRES \\
2455314.122   &   -5.90   &   1.86   &   HIRES \\
2455399.913   &   -2.46   &   1.98   &   HIRES \\
2455410.786   &   -18.38   &   2.07   &   HIRES \\
2455468.760   &   -13.46   &   1.95   &   HIRES \\
2455486.748   &   -7.26   &   2.01   &   HIRES \\
2455636.121   &   -4.12   &   2.13   &   HIRES \\
2455749.987   &   1.82   &   2.27   &   HIRES \\
2455749.994   &   1.72   &   2.32   &   HIRES \\
2455824.890   &   3.22   &   2.16   &   HIRES \\
2455839.808   &   1.25   &   1.95   &   HIRES \\
2455839.815   &   -0.84   &   2.07   &   HIRES \\
2456027.129   &   3.54   &   2.10   &   HIRES \\
2456116.914   &   1.40   &   1.86   &   HIRES \\
2456116.921   &   3.80   &   1.80   &   HIRES \\
2456117.872   &   6.44   &   1.86   &   HIRES \\
2456117.880   &   5.28   &   1.86   &   HIRES \\
2456141.994   &   -2.49   &   1.83   &   HIRES \\
2456142.001   &   -5.05   &   1.98   &   HIRES \\
2456168.851   &   2.38   &   2.10   &   HIRES \\
2456168.859   &   4.33   &   1.92   &   HIRES \\
2456169.809   &   10.32   &   2.24   &   HIRES \\
2456169.816   &   8.56   &   2.27   &   HIRES \\
2456385.894   &   2.59   &   0.92   &   HARPS-pre \\
2456386.905   &   3.40   &   0.77   &   HARPS-pre \\
2456387.890   &   0.54   &   0.61   &   HARPS-pre \\
2456388.912   &   1.00   &   0.90   &   HARPS-pre \\
2456389.883   &   4.79   &   0.79   &   HARPS-pre \\
2456390.909   &   2.57   &   1.06   &   HARPS-pre \\
2456391.869   &   1.74   &   0.85   &   HARPS-pre \\
2456393.909   &   4.97   &   0.74   &   HARPS-pre \\
2456395.913   &   1.43   &   0.68   &   HARPS-pre \\
2456397.901   &   1.28   &   0.79   &   HARPS-pre \\
2456398.899   &   0.65   &   0.77   &   HARPS-pre \\
2456399.895   &   2.44   &   0.87   &   HARPS-pre \\
2456400.878   &   -0.19   &   0.79   &   HARPS-pre \\
  
\hline           
\end{tabular}
\end{table}

\begin{table}
\caption*{Doppler measurements for HD\,180617 -- continued.}
\label{table:V1428_RVs_d} 

\centering  

\begin{tabular}{c c c l } 

\hline\hline    
\noalign{\vskip 0.5mm}

JD & RV [m\,s$^{-1}$] & $\sigma_{RV}$ [m\,s$^{-1}$]   & Instrument \\   

\hline     
\noalign{\vskip 0.5mm}    

2456402.840   &   2.36   &   0.74   &   HARPS-pre \\
2456402.928   &   0.36   &   0.74   &   HARPS-pre \\
2456403.931   &   0.77   &   0.74   &   HARPS-pre \\
2456404.916   &   0.44   &   0.61   &   HARPS-pre \\
2456405.931   &   -0.76   &   1.04   &   HARPS-pre \\
2456406.914   &   -0.70   &   0.95   &   HARPS-pre \\
2456407.876   &   -1.74   &   0.79   &   HARPS-pre \\
2456409.915   &   1.45   &   0.85   &   HARPS-pre \\
2456411.879   &   -1.09   &   0.74   &   HARPS-pre \\
2456415.828   &   0.98   &   1.08   &   HARPS-pre \\
2456415.924   &   1.41   &   0.79   &   HARPS-pre \\
2456416.893   &   -0.73   &   0.79   &   HARPS-pre \\
2456433.019   &   1.21   &   2.32   &   HIRES \\
2456433.026   &   3.67   &   2.29   &   HIRES \\
2456434.056   &   -8.11   &   2.13   &   HIRES \\
2456434.062   &   -4.27   &   2.01   &   HIRES \\
2456434.883   &   -3.51   &   0.77   &   HARPS-pre \\
2456435.902   &   -4.57   &   0.79   &   HARPS-pre \\
2456437.867   &   -2.49   &   0.61   &   HARPS-pre \\
2456452.840   &   -4.50   &   0.71   &   HARPS-pre \\
2456453.834   &   -3.89   &   7.15   &   HARPS-pre \\
2456455.770   &   -5.29   &   1.12   &   HARPS-pre \\
2456456.833   &   -5.23   &   0.85   &   HARPS-pre \\
2456469.762   &   -1.01   &   0.58   &   HARPS-pre \\
2456471.831   &   3.05   &   0.97   &   HARPS-pre \\
2456472.789   &   -0.96   &   0.74   &   HARPS-pre \\
2456474.819   &   1.67   &   0.74   &   HARPS-pre \\
2456475.731   &   -1.77   &   0.77   &   HARPS-pre \\
2456477.727   &   0.21   &   0.71   &   HARPS-pre \\
2456498.757   &   4.64   &   0.74   &   HARPS-pre \\
2456505.641   &   4.80   &   0.77   &   HARPS-pre \\
2456521.596   &   -0.28   &   0.74   &   HARPS-pre \\
2456524.611   &   0.92   &   0.54   &   HARPS-pre \\
2456525.652   &   0.67   &   0.74   &   HARPS-pre \\
2456526.683   &   0.36   &   0.71   &   HARPS-pre \\
2456538.602   &   -3.73   &   0.87   &   HARPS-pre \\
2456539.637   &   -4.41   &   0.77   &   HARPS-pre \\
2456541.604   &   -4.07   &   0.58   &   HARPS-pre \\
2456543.580   &   -1.58   &   0.77   &   HARPS-pre \\
2456548.803   &   1.35   &   1.86   &   HIRES \\
2456548.809   &   -6.37   &   2.10   &   HIRES \\
2456549.765   &   -6.53   &   2.16   &   HIRES \\
2456549.771   &   -5.63   &   1.95   &   HIRES \\
2456551.804   &   2.99   &   2.13   &   HIRES \\
2456551.811   &   5.16   &   2.10   &   HIRES \\
2456585.539   &   -4.75   &   0.61   &   HARPS-pre \\
2456586.508   &   -4.05   &   0.68   &   HARPS-pre \\
2456764.845   &   -3.59   &   0.61   &   HARPS-pre \\
2456767.853   &   -4.59   &   0.68   &   HARPS-pre \\
2456814.830   &   0.97   &   0.61   &   HARPS-pre \\
2456815.845   &   3.08   &   0.77   &   HARPS-pre \\
2456816.830   &   3.44   &   0.64   &   HARPS-pre \\
2456817.777   &   2.45   &   0.82   &   HARPS-pre \\
2456819.838   &   -0.73   &   0.74   &   HARPS-pre \\
2456822.838   &   -0.61   &   1.44   &   HARPS-pre \\
2456824.842   &   -4.06   &   0.71   &   HARPS-pre \\
2456837.850   &   3.75   &   0.71   &   HARPS-pre \\
2456838.821   &   4.12   &   0.54   &   HARPS-pre \\
2456839.803   &   4.74   &   0.64   &   HARPS-pre \\
2456840.801   &   5.97   &   0.58   &   HARPS-pre \\
  
\hline           
\end{tabular}
\end{table}

\clearpage

\begin{table}
\caption*{Doppler measurements for HD\,180617 -- continued.}
\label{table:V1428_RVs_e} 

\centering  

\begin{tabular}{c c c l } 

\hline\hline    
\noalign{\vskip 0.5mm}

JD & RV [m\,s$^{-1}$] & $\sigma_{RV}$ [m\,s$^{-1}$]   & Instrument \\  

\hline     
\noalign{\vskip 0.5mm}    

2456857.790   &   -1.33   &   0.77   &   HARPS-pre \\
2456858.768   &   -0.60   &   0.71   &   HARPS-pre \\
2456862.743   &   -0.06   &   0.64   &   HARPS-pre \\
2456863.712   &   -1.95   &   0.68   &   HARPS-pre \\
2456864.717   &   -2.77   &   0.61   &   HARPS-pre \\
2456871.656   &   -4.51   &   0.77   &   HARPS-pre \\
2456872.710   &   -4.64   &   0.71   &   HARPS-pre \\
2456910.824   &   10.64   &   1.98   &   HIRES \\
2456915.636   &   2.67   &   0.90   &   HARPS-pre \\
2456916.543   &   1.23   &   0.77   &   HARPS-pre \\
2456919.632   &   2.65   &   0.95   &   HARPS-pre \\
2456920.602   &   3.64   &   0.82   &   HARPS-pre \\
2456922.607   &   0.74   &   0.74   &   HARPS-pre \\
2456923.594   &   3.22   &   0.87   &   HARPS-pre \\
2456924.606   &   1.92   &   0.74   &   HARPS-pre \\
2456925.568   &   2.73   &   0.58   &   HARPS-pre \\
2456926.592   &   2.82   &   0.61   &   HARPS-pre \\
2456927.524   &   7.59   &   1.06   &   HARPS-pre \\
2456929.565   &   4.02   &   0.54   &   HARPS-pre \\
2456930.568   &   4.68   &   0.68   &   HARPS-pre \\
2456931.568   &   3.93   &   0.71   &   HARPS-pre \\
2456936.536   &   7.70   &   0.79   &   HARPS-pre \\
2457185.869   &   -7.35   &   1.00   &    HARPS-post \\
2457186.847   &   -5.82   &   0.95   &    HARPS-post \\
2457187.839   &   -6.33   &   0.89   &    HARPS-post \\
2457188.856   &   -8.29   &   0.88   &    HARPS-post \\
2457189.846   &   -5.11   &   0.99   &    HARPS-post \\
2457190.841   &   -6.56   &   0.95   &    HARPS-post \\
2457204.683   &   -10.83   &   0.99   &    HARPS-post \\
2457204.832   &   -10.45   &   0.86   &    HARPS-post \\
2457210.818   &   -4.06   &   1.13   &    HARPS-post \\
2457214.786   &   -6.71   &   1.05   &    HARPS-post \\
2457249.701   &   -3.29   &   1.45   &    HARPS-post \\
2457252.592   &   -4.14   &   1.17   &    HARPS-post \\
2457253.596   &   -4.47   &   1.00   &    HARPS-post \\
2457256.536   &   -5.82   &   1.09   &    HARPS-post \\
2457257.639   &   -4.50   &   1.05   &    HARPS-post \\
2457259.582   &   -2.82   &   1.39   &    HARPS-post \\
2457260.543   &   -2.75   &   1.39   &    HARPS-post \\
2457263.646   &   -3.59   &   1.21   &    HARPS-post \\
2457264.648   &   -5.42   &   1.02   &    HARPS-post \\
2457265.643   &   -3.81   &   1.07   &    HARPS-post \\
2457266.631   &   -4.89   &   1.36   &    HARPS-post \\
2457267.666   &   -3.90   &   1.21   &    HARPS-post \\
2457269.664   &   -4.95   &   1.30   &    HARPS-post \\
2457270.624   &   -3.54   &   1.36   &    HARPS-post \\
2457277.540   &   -3.64   &   1.20   &    HARPS-post \\
2457292.568   &   -9.16   &   1.20   &    HARPS-post \\
2457294.552   &   -7.15   &   1.36   &    HARPS-post \\
2457472.899   &   -2.33   &   0.88   &    HARPS-post \\
2457477.715   &   -2.67   &   1.20   &   CARMENES \\
2457490.701   &   -2.54   &   1.50   &   CARMENES \\
2457495.692   &   -4.44   &   2.05   &   CARMENES \\
2457504.864   &   -7.80   &   1.39   &    HARPS-post \\
2457510.697   &   -4.92   &   1.16   &   CARMENES \\
2457534.675   &   -4.58   &   1.59   &   CARMENES \\
2457537.655   &   -6.54   &   1.79   &   CARMENES \\
2457539.599   &   -1.05   &   1.59   &   CARMENES \\
2457541.678   &   -7.37   &   1.77   &   CARMENES \\
2457544.671   &   -2.22   &   1.50   &   CARMENES \\
  
\hline           
\end{tabular}
\end{table}

\begin{table}
\caption*{Doppler measurements for HD\,180617 -- continued.}
\label{table:V1428_RVs_f} 

\centering  

\begin{tabular}{c c c l } 

\hline\hline    
\noalign{\vskip 0.5mm}

JD & RV [m\,s$^{-1}$] & $\sigma_{RV}$ [m\,s$^{-1}$]   & Instrument \\   

\hline     
\noalign{\vskip 0.5mm}    

2457553.647   &   2.62   &   1.28   &   CARMENES \\
2457555.597   &   4.90   &   1.57   &   CARMENES \\
2457596.408   &   -1.48   &   1.87   &   CARMENES \\
2457597.484   &   -3.80   &   1.86   &   CARMENES \\
2457612.677   &   -6.76   &   1.09   &    HARPS-post \\
2457613.643   &   -4.07   &   1.21   &    HARPS-post \\
2457614.672   &   -4.47   &   1.13   &    HARPS-post \\
2457615.633   &   -5.24   &   0.69   &    HARPS-post \\
2457622.509   &   -3.25   &   2.17   &   CARMENES \\
2457626.418   &   -4.67   &   1.87   &   CARMENES \\
2457629.402   &   -1.37   &   2.07   &   CARMENES \\
2457629.653   &   -7.39   &   1.27   &    HARPS-post \\
2457633.354   &   1.89   &   2.23   &   CARMENES \\
2457636.497   &   -1.63   &   1.82   &   CARMENES \\
2457644.393   &   -2.32   &   1.41   &   CARMENES \\
2457646.364   &   -1.60   &   1.47   &   CARMENES \\
2457647.319   &   7.42   &   3.16   &   CARMENES \\
2457652.363   &   4.77   &   2.60   &   CARMENES \\
2457655.352   &   4.94   &   2.42   &   CARMENES \\
2457657.411   &   3.57   &   2.21   &   CARMENES \\
2457678.288   &   1.41   &   1.88   &   CARMENES \\
2457679.283   &   0.18   &   2.14   &   CARMENES \\
2457694.301   &   -0.26   &   1.41   &   CARMENES \\
2457704.308   &   0.18   &   1.42   &   CARMENES \\
2457823.701   &   -3.91   &   1.42   &   CARMENES \\
2457828.679   &   -0.46   &   1.31   &   CARMENES \\
2457834.711   &   -2.54   &   1.42   &   CARMENES \\
2457848.633   &   -3.70   &   1.42   &   CARMENES \\
2457855.699   &   -1.38   &   1.16   &   CARMENES \\
2457856.692   &   -1.52   &   1.30   &   CARMENES \\
2457858.647   &   0.23   &   1.21   &   CARMENES \\
2457859.679   &   2.27   &   1.35   &   CARMENES \\
2457864.648   &   3.31   &   1.39   &   CARMENES \\
2457866.677   &   4.54   &   1.73   &   CARMENES \\
2457877.619   &   -0.28   &   1.33   &   CARMENES \\
2457879.587   &   -6.25   &   2.39   &   CARMENES \\
2457887.592   &   -1.03   &   1.39   &   CARMENES \\
2457889.665   &   -1.04   &   1.36   &   CARMENES \\
2457892.607   &   0.25   &   1.36   &   CARMENES \\
2457894.610   &   -0.23   &   1.38   &   CARMENES \\
2457898.581   &   -2.58   &   1.28   &   CARMENES \\
2457905.533   &   -0.52   &   1.36   &   CARMENES \\
2457907.558   &   0.35   &   1.30   &   CARMENES \\
2457912.590   &   -0.70   &   1.59   &   CARMENES \\
2457914.533   &   -0.04   &   1.55   &   CARMENES \\
2457915.593   &   -1.95   &   1.35   &   CARMENES \\
2457918.667   &   -4.14   &   1.82   &   CARMENES \\
2457920.525   &   -5.20   &   1.50   &   CARMENES \\
2457922.538   &   -3.06   &   1.14   &   CARMENES \\
2457934.452   &   -3.15   &   2.74   &   CARMENES \\
2457935.475   &   -1.55   &   1.20   &   CARMENES \\
2457942.487   &   1.61   &   1.35   &   CARMENES \\
2457943.570   &   -3.44   &   1.57   &   CARMENES \\
2457946.576   &   -3.36   &   1.38   &   CARMENES \\
2457947.455   &   -2.82   &   1.62   &   CARMENES \\
2457948.501   &   -2.51   &   1.53   &   CARMENES \\
2457949.501   &   -1.09   &   1.21   &   CARMENES \\
2457950.464   &   1.07   &   1.44   &   CARMENES \\
2457951.531   &   -2.77   &   1.50   &   CARMENES \\
2457953.611   &   2.10   &   1.65   &   CARMENES \\
  
\hline           
\end{tabular}
\end{table}

\clearpage

\begin{table}
\caption*{Doppler measurements for HD\,180617 -- continued.} 
\label{table:V1428_RVs_g} 

\centering  

\begin{tabular}{c c c l } 

\hline\hline    
\noalign{\vskip 0.5mm}

JD & RV [m\,s$^{-1}$] & $\sigma_{RV}$ [m\,s$^{-1}$]   & Instrument \\ 

\hline     
\noalign{\vskip 0.5mm}    

2457954.493   &   -0.37   &   1.44   &   CARMENES \\
2457955.616   &   2.48   &   1.42   &   CARMENES \\
2457956.620   &   -0.52   &   1.59   &   CARMENES \\
2457958.538   &   5.36   &   1.47   &   CARMENES \\
2457959.597   &   2.51   &   1.42   &   CARMENES \\
2457960.486   &   4.19   &   1.23   &   CARMENES \\
2457961.426   &   3.61   &   1.21   &   CARMENES \\
2457962.454   &   2.91   &   1.25   &   CARMENES \\
2457963.469   &   3.65   &   1.21   &   CARMENES \\
2457964.464   &   2.20   &   1.25   &   CARMENES \\
2457965.457   &   2.62   &   1.39   &   CARMENES \\
2457968.487   &   2.70   &   1.39   &   CARMENES \\
2457969.488   &   3.75   &   1.30   &   CARMENES \\
2457970.569   &   3.48   &   1.26   &   CARMENES \\
2457971.441   &   2.20   &   1.35   &   CARMENES \\
2457975.436   &   4.70   &   1.41   &   CARMENES \\
2457976.449   &   2.42   &   1.69   &   CARMENES \\
2457977.430   &   4.25   &   1.55   &   CARMENES \\
2457979.419   &   4.01   &   1.39   &   CARMENES \\
2457982.464   &   4.63   &   1.39   &   CARMENES \\
2457985.423   &   4.56   &   1.60   &   CARMENES \\
2457987.430   &   1.81   &   1.21   &   CARMENES \\
2457989.373   &   -0.72   &   1.88   &   CARMENES \\
2457990.439   &   -1.64   &   1.44   &   CARMENES \\
2457997.376   &   -0.88   &   1.30   &   CARMENES \\
2457998.433   &   -3.80   &   3.47   &   CARMENES \\
2457999.325   &   3.91   &   1.42   &   CARMENES \\
2458000.417   &   0.25   &   1.09   &   CARMENES \\
2458001.366   &   0.47   &   1.18   &   CARMENES \\
2458002.358   &   2.35   &   1.25   &   CARMENES \\
2458005.324   &   -0.92   &   1.14   &   CARMENES \\
2458006.348   &   1.44   &   1.68   &   CARMENES \\
2458007.351   &   0.51   &   1.18   &   CARMENES \\
2458009.353   &   1.90   &   1.26   &   CARMENES \\
2458010.348   &   0.24   &   1.12   &   CARMENES \\
2458017.364   &   -1.62   &   1.35   &   CARMENES \\
2458022.340   &   0.09   &   1.42   &   CARMENES \\
2458023.373   &   -2.11   &   1.35   &   CARMENES \\
2458027.365   &   -1.00   &   1.39   &   CARMENES \\
2458030.388   &   -3.43   &   1.70   &   CARMENES \\
2458031.396   &   -2.35   &   1.86   &   CARMENES \\
2458035.374   &   -3.11   &   2.49   &   CARMENES \\
2458042.369   &   6.02   &   3.61   &   CARMENES \\
2458047.350   &   -5.59   &   1.70   &   CARMENES \\
2458048.397   &   -3.19   &   1.52   &   CARMENES \\
2458053.350   &   -3.85   &   1.56   &   CARMENES \\
2458054.351   &   -2.81   &   1.42   &   CARMENES \\
2458055.267   &   -2.93   &   1.20   &   CARMENES \\
2458056.257   &   -2.24   &   1.45   &   CARMENES \\
2458057.247   &   -1.05   &   1.21   &   CARMENES \\
2458058.266   &   -0.23   &   1.44   &   CARMENES \\
2458059.297   &   -0.26   &   1.21   &   CARMENES \\
2458060.264   &   -3.90   &   1.25   &   CARMENES \\
2458065.250   &   0.44   &   1.36   &   CARMENES \\
2458066.253   &   -3.99   &   1.52   &   CARMENES \\
2458080.284   &   0.14   &   1.23   &   CARMENES \\
2458081.265   &   0.64   &   1.44   &   CARMENES \\
2458082.276   &   2.14   &   1.55   &   CARMENES \\
2458096.235   &   -0.16   &   1.90   &   CARMENES \\
2458099.236   &   9.41   &   4.58   &   CARMENES \\ 
  
\hline           
\end{tabular}
\end{table}

\end{appendix}

\end{document}